\documentclass[rmp,amsfonts,showpacs,showkeys]{revtex4}
\usepackage[T1]{fontenc}
\usepackage{graphicx}
\RequirePackage{times}
\RequirePackage{mathptm}

\def\frak{\bf}
\def\Bbb{\bf}
 \def\sp{\textrm{Sp}}
\def\hh{{\frak H}}
\def\ll{{\frak C (\frak H)}}
\def\mm{{\frak M}}
\def\ppp{{\frak P} (\frak H)}
\begin{document}

\title{Quantum logic. A brief outline}
\author{Karl Svozil}
\email{svozil@tuwien.ac.at}
\homepage{http://tph.tuwien.ac.at/~svozil}
\affiliation{Institut f\"ur Theoretische Physik, University of Technology Vienna,
Wiedner Hauptstra\ss e 8-10/136, A-1040 Vienna, Austria}

\begin{abstract}
Quantum logic has been introduced by Birkhoff and von Neumann as an attempt to base the logical primitives, the propositions and the relations and operations among them, on quantum theoretical entities, and thus on the related empirical evidence of the quantum world. We give a brief outline of quantum logic, and some of its algebraic properties, such as nondistributivity, whereby emphasis is given to concrete experimental setups related to quantum logical entities.
A probability theory based on quantum logic is fundamentally and sometimes even spectacularly different from probabilities based on classical Boolean logic. We give a brief outline of its nonclassical aspects; in particular violations of Boole-Bell type consistency constraints on joint probabilities, as well as the Kochen-Specker theorem, demonstrating in a constructive, finite way the scarcity and even nonexistence of two-valued states interpretable as classical truth assignments.
{A more complete introduction of the author can be found in the book {\it Quantum Logic (Springer, 1998)}}
\end{abstract}

\maketitle

\section{Basic ideas}
Quantum logic has been introduced by
Garrett Birkhoff and John von Neumann in the thirties
\cite{birkhoff-36}.
They  organized it {\em top-down}:
The starting point is von Neumann's Hilbert space
formalism of quantum mechanics.
In a second step, certain entities of Hilbert spaces are identified
with propositions,
partial order relations and lattice operations --- Birkhoff's field of
expertise.
These relations and operations can then be
associated with the logical
implication relation and operations such as {\it and}, {\it or}, and
{\it not}. Thereby, a
 ``nonclassical,'' nonboolean
logical structure is induced which originates in theoretical
physics.
If theoretical physics is taken as a faithful
representation of our experience, such an ``operational''
\cite{bridgman27,bridgman,bridgman52}
logic derives
its justification by the phenomena themselves.
In this sense,
one of the main ideas behind quantum logic is the quasi-inductive
construction of the logical and algebraic order of events from empirical
facts.

This is very different from the ``classical''  logical approach, which
is also {\em top-down}: There, the system of symbols, the axioms, as
well as the rules of inference are mentally constructed,
abstract objects of
our thought.
Insofar as our thoughts can pretend to exist independent from
the physical Universe, such ``classical'' logical systems can be
conceived as totally independent from the world of the
phenomena.

In the following we shall shortly review quantum logic. More
detailed introduction can be found in the books of
 Mackey
\cite{ma-57},
 Jauch
\cite{jauch},
Varadarajan \cite{varadarajanI,varadarajanII},
Piron
\cite{piron-76},
Marlow
\cite{marlow},
Gudder
\cite{gudder:79,gudder},  Maczy\'nski \cite{maczy},
Beltrametti and  Cassinelli
\cite{bell-cas},
Kalmbach
\cite{kalmbach-83,kalmbach-86},
Cohen
\cite{cohen},
Pt{\'{a}}k and Pulmannov{\'{a}}
\cite{pulmannova-91},
Giuntini
\cite{giuntini-91}, and in a forthcoming book of the author
\cite{svozil-ql},
among
others. A  bibliography on quantum logics and related
structures has been compiled by
 Pavi{\v{c}}i{\'{c}} \cite{pavicic-92}.

\begin{description}
\item[$\bullet$]
Any closed linear
subspace of --- or, equivalently, any
projection operator on --- a Hilbert space corresponds to an elementary
proposition. The elementary {\it true}--{\it false} proposition can in
English be spelled out explicitly as
\begin{quote}
``The physical system has a property corresponding to the associated
closed linear subspace.''
\end{quote}

\item[$\bullet$]
The logical {\it and} operation is identified with the set
theoretical intersection of two propositions ``$\cap$''; i.e., with the
intersection of two subspaces.
It is denoted by the symbol ``$\wedge$''.
So, for two
propositions $p$ and $q$ and their associated closed linear
subspaces
${\frak M}_p$ and
${\frak M}_q$,
$$
{\frak M}_{p\wedge q} = \{x \mid x\in
{\frak M}_p, \;
x\in {\frak M}_q\} .$$

\item[$\bullet$]
The logical {\it or} operation is identified with the closure of the
linear span ``$\oplus$'' of the subspaces corresponding to the two
propositions.\footnote{
Notice that
a vector of Hilbert space may be an element of
$
{\frak M}_{p} \oplus
{\frak M}_{q}
$
without being an element of either
$
{\frak M}_{p} $ or
${\frak M}_{q}
$, since
$
{\frak M}_{p} \oplus
{\frak M}_{q}
$
includes all the vectors in
$
{\frak M}_{p} \cup
{\frak M}_{q}
$, as well as all of their linear combinations (superpositions) and
their limit vectors.
}
 It is denoted by the symbol ``$\vee$''.
So, for two
propositions $p$ and $q$ and their associated closed linear
subspaces
${\frak M}_p$ and
${\frak M}_q$,
$$
{\frak M}_{p\vee q} =
{\frak M}_{p} \oplus
{\frak M}_{q} =
 \{x \mid x=\alpha x+\beta z,\; \alpha,\beta \in {\Bbb C},\; y\in
{\frak M}_p, \;
z\in {\frak M}_q\} .$$

The symbol $\oplus$ will used to indicate the closed linear subspace
spanned by two vectors. That is,
$$u\oplus v=\{ w\mid w=\alpha u+ \beta v,\; \alpha,\beta \in {\Bbb C}
,\; u,v \in {\frak H}\}.$$
More generally, the symbol $\oplus$  indicates the closed linear
subspace spanned by two linear subspaces.
That is, if $u,v\in {\frak C}({{\frak H}})$, where ${\frak
C}({{\frak H}})$
stands for the set of all subspaces of the Hilbert space, then
$$u\oplus v=\{ w\mid w=\alpha u+ \beta v,\; \alpha,\beta \in {\Bbb R}
,\; u,v \in {\frak C}({{\frak H}}) \}.$$

\item[$\bullet$]
The logical {\it not}-operation, or the ``complement'',
is
identified with operation of taking the orthogonal subspace ``$\perp$''.
It is denoted by the symbol ``~$'$~''.
In particular, for a
proposition $p$ and its associated closed linear
subspace
${\frak M}_p$,
$$
{\frak M}_{p'} =
 \{x \mid (x,y)=0,\; y\in
{\frak M}_p
\} .$$

\item[$\bullet$]
The logical {\it implication} relation is identified with the set
theoretical subset relation ``$\subset$''.
It is denoted by the symbol ``$\rightarrow$''.
So, for two
propositions $p$ and $q$ and their associated closed linear
subspaces
${\frak M}_p$ and
${\frak M}_q$,
$$
{p\rightarrow q} \Longleftrightarrow
{\frak M}_{p} \subset
{\frak M}_{q}.$$

\item[$\bullet$]
A trivial statement which is always {\it true} is denoted by $1$.
It is represented by the entire Hilbert space $\frak H$.
So, $${\frak M}_1=\frak H.$$

\item[$\bullet$]
An absurd statement which is always {\it false} is denoted by $0$.
It is represented by the zero vector $0$.
So, $${\frak M}_0= 0.$$
\end{description}
Let us verify some logical statements.
\begin{description}
\item[$(p')'=p$:]
For closed orthogonal subspaces of ${\frak H}$,
$
{\frak M}_{(p')'}
=
\{x\mid (x,y)=0, y\in
\{z\mid (z,u)=0, u\in
{\frak M}_{p} \} \} =
{\frak M}_{p}.
$
\item[$1'=0$:]
$
{\frak M}_{1'}
=
\{x\mid (x,y)=0, y\in {\frak H}\} ={\frak M}_0= 0.
$
\item[$0'=1$:]
$
{\frak M}_{0'}
=
\{x\mid (x,y)=0, y=0\} ={\frak M}_1= {\frak H}.
$

\item[$p\vee p'=1$:]
$
{\frak M}_{p\vee p'}=
{\frak M}_{p} \oplus
{\frak M}_{p'}
=
 \{x \mid x=\alpha y+\beta z,\; \alpha,\beta \in {\Bbb C},\; y\in
{\frak M}_p, \;
z\in {\frak M}_{p'}\} ={\frak M}_1 .$

\item[$p\wedge p'=0$:]
$
{\frak M}_{p\wedge p'}=
{\frak M}_{p} \cap
{\frak M}_{p'}
=
 \{x \mid x\in
{\frak M}_p, \;
x\in {\frak M}_{p'}\} ={\frak M}_0 .$

\end{description}

Table
 \ref{tcompa} lists the identifications of relations of operations of
various lattice types.
\begin{table}[h]
\begin{center}
{\footnotesize
 \begin{tabular}{ccccc} \hline\hline
 generic lattice  &  order relation $\rightarrow $  & ``meet'' $\sqcap$
&
``join''
$\sqcup$ & ``complement'' $'$\\
\hline
``classical'' lattice  &  subset $\subset $  & intersection $\cap$ &
union
$\cup$ & complement\\
of subsets&&&&\\
of a set&&&&\\
\hline
propositional&implication&disjunction&conjunction&negation\\
calculus&$\rightarrow$&``${\it and}$'' $\wedge$&``${\it or}$''
$\vee$&``${\it not}$''$\neg$\\
\hline
Hilbert & subspace& intersection of & closure of     & orthogonal \\
lattice & relation& subspaces $\cap$&  linear& subspace   \\
        & $\subset$ &                 & span $\oplus$  &  $\perp$   \\
\hline
lattice of& $E_1E_2=E_1$& $E_1E_2$&
                               $E_1+E_2-E_1E_2$& orthogonal\\
commuting&&($\lim_{n\rightarrow \infty}(E_1E_2)^n$)&&projection\\
(noncommuting)\\
projection\\
operators\\
 \hline\hline
 \end{tabular}
}
 \caption{Comparison of the identifications of lattice relations and
 operations for the lattices of subsets of a set, for
 experimental propositional calculi, for  Hilbert lattices, and for
lattices of commuting projection operators.
 \label{tcompa}}
 \end{center} \end{table}

Propositional structures are often represented by Hasse and Greechie
diagrams.
A Hasse diagram is a convenient representation of the
logical implication,
as well as of the {\it and} and {\it or}
operations
among propositions.
 Points
``~$\bullet$~'' represent propositions. Propositions
which are implied by other ones are drawn higher than the other ones.
Two propositions are connected by a line if one implies the other.

A much more compact representation of the propositional calculus can be
given in terms of
its Greechie diagram.
 There, the  points ``~$\circ$~'' represent
the atoms.
\label{greechie-diagram}
\index{Greechie diagram}
If they belong to the same Boolean
algebra, they are connected by edges or smooth curves.
We will later use ``almost''
Greechie diagrams,
omitting points which belong to only one curve. This makes the
diagrams a bit more comprehensive.

\section{Complementarity}
Let us call two propositions $p,q$ {\em comeasurable}
or {\em compatible}
if and only if there exist mutually orthogonal propositions  $a,b,c$
such that $p=a\vee b$ and $q=a\vee c$
\cite[p. 118]{varadarajanI}.
Intuitively, we may assume that two comeasurable propositions $c\vee a$
and $c\vee b$ consist of an
``identical'' part $a$, as well as of the orthogonal  parts $b,c$ (which
are also orthogonal to $a$).

  Clearly, orthogonality implies
comeasurability, since if
$p$ and
$q$ are orthogonal, we may identify
$a, b, c$ with $0,p,q$, respectively.

If one is willing to give meaning to noncomeasurable blocks of
observables and thus to counterfactuals, it is straightforward to
proceed with the formalism.

Consider a collection of blocks.
 Some of these blocks may have a common nontrivial
observable. The complete logic with respect to the collection of the
blocks is obtained by the following construction.
\index{pasting}
\begin{description}
\item[$\bullet$]
The tautologies of all blocks are identified.
\item[$\bullet$]
The absurdities of all blocks are identified.
\item[$\bullet$]
Identical elements in different blocks are identified.
\item[$\bullet$]
The logical and algebraical structures of all blocks remain intact.
\end{description}
This construction is often referred to as {\em pasting} construction.
If the blocks are only pasted together at the tautology and
the absurdity, one calls the resulting logic a {\em horizontal
sum}.\footnote{This definition of ``horizontal sum'' is equivalent to
the coproduct of complemented lattices. A coproduct is like a least
upper bound: each component maps onto it, and if each component maps
into another complemented lattice, then the coproduct also does in a
unique way.} \label{l-horsum}
\index{horizontal sum}
In a sense, the pasting construction allows one to obtain a {\em global}
representation of different universes which are defined (and classical)
{\em locally}. This local --- {\it versus} --- global theme will be
discussed
below.
A pasting of two propositional structures $L_1$ and $L_2$ will be
denoted by
$$L_1\oplus L_2.$$
 For further reading, see
also
Gudrun Kalmbach \cite{kalmbach-83,kalmbach-86},
Robert Piziak
\cite{piz-88},
Pavel Pt{\'{a}}k and Sylvia Pulmannov{\'{a}}
\cite{pulmannova-91}, and
Mirko Navara and V. Rogalewicz
\cite{nav:91}.

\section{Spin one-half}
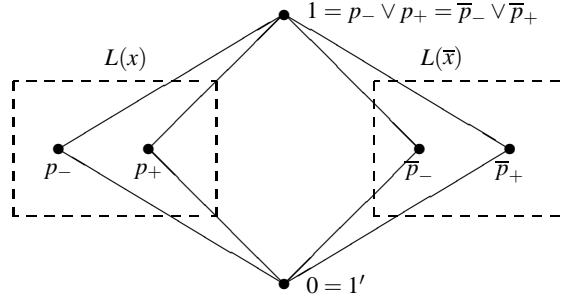
\begin{figure}[htd]
\begin{center}
\unitlength 0.60mm
\linethickness{0.4pt}
\begin{picture}(125.00,60.73)
\put(60.00,0.00){\circle*{2.11}}
\put(30.00,30.00){\circle*{2.11}}
\put(60.00,59.67){\circle*{2.11}}
\put(90.00,30.00){\circle*{2.11}}
\put(60.00,0.00){\line(-1,1){30.00}}
\put(30.00,30.00){\line(1,1){30.00}}
\put(60.00,60.00){\line(1,-1){30.00}}
\put(90.00,30.00){\line(-1,-1){30.00}}
\put(65.00,0.00){\makebox(0,0)[lc]{$0=1'$}}
\put(65.00,60.00){\makebox(0,0)[lc]{$1=p_-\vee p_+=\overline p_-\vee\overline p_+$}}
\put(30.00,25.00){\makebox(0,0)[cc]{$p_+$}}
\put(90.00,25.00){\makebox(0,0)[cc]{$\overline p_-$}}
\put(60.00,0.00){\line(-5,3){50.00}}
\put(10.00,30.00){\line(5,3){50.00}}
\put(60.00,60.00){\line(5,-3){50.00}}
\put(110.00,30.00){\line(-5,-3){50.00}}
\put(10.00,30.00){\circle*{2.11}}
\put(10.00,25.00){\makebox(0,0)[cc]{$p_-$}}
\put(110.00,30.00){\circle*{2.11}}
\put(110.00,25.00){\makebox(0,0)[cc]{$\overline p_+$}}
\put(25.00,50.00){\makebox(0,0)[cc]{$L(x)$}}
\put(95.00,50.00){\makebox(0,0)[cc]{$L(\overline x)$}}
\put(0.00,15.00){\dashbox{2.00}(45.00,30.00)[cc]{}}
\put(80.00,15.00){\dashbox{2.00}(45.00,30.00)[cc]{}}
\end{picture}
\end{center}
\caption{\label{f-hd-mo2}
Hasse diagram of the ``Chinese lantern'' form obtained by the pasting of
two spin one-half state propositional systems
$L(x)$ and
$L(\overline x)$ which are noncomeasurable.
The resulting logical structure is a modular orthocomplemented lattice
$L(x)\oplus L(\overline{x}) = MO_2$.
\index{$MO_2$}
The blocks (without $0,1$) are indicated by dashed boxes. They will be
henceforth omitted.
}
\end{figure}

\begin{figure}[htd]
\begin{center}
\unitlength 0.50mm
\linethickness{0.4pt}
\begin{picture}(131.06,15.00)
\put(0.00,5.00){\circle{2.11}}
\put(60.00,5.00){\circle{2.11}}
\put(0.00,0.00){\makebox(0,0)[cc]{$p_-$}}
\put(60.00,0.00){\makebox(0,0)[cc]{$p_+$}}
\put(0.00,5.00){\line(1,0){60.00}}
\put(70.00,5.00){\circle{2.11}}
\put(130.00,5.00){\circle{2.11}}
\put(70.00,0.00){\makebox(0,0)[cc]{$\overline p_-$}}
\put(130.00,0.00){\makebox(0,0)[cc]{$\overline p_+$}}
\put(70.00,5.00){\line(1,0){60.00}}
\put(30.00,15.00){\makebox(0,0)[cc]{$L(x)$}}
\put(100.00,15.00){\makebox(0,0)[cc]{$L(\overline x)$}}
\end{picture}
\end{center}
\caption{\label{f-gd-mo2}
Greechie
diagram of two spin one-half state propositional systems $L(x)$ and
$L(\overline x)$ which are noncomeasurable.
}
\end{figure}
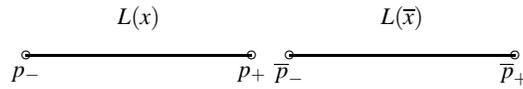

As a first example, we shall paste together observables of the spin
one-half systems.
We have associated a propositional system
$$L(x)= \{ 0, p_{-}, p_{+}, 1 \}, $$
corresponding to the outcomes of a measurement of the spin states
along the $x$-axis (any other direction would have done as well).
If the spin states would be measured along a different spatial
direction, say
$\overline{x}\neq x \; \textrm{mod} \pi$, an identical propositional
system
$$L(\overline x)= \{ \overline 0, \overline p_{-}, \overline p_{+},
\overline 1 \} $$
would have been the result.
$L(x)$ and
$L(\overline x)$ can be jointly represented by pasting them together.
In particular, we identify their tautologies and absurdities; i.e.,
\begin{eqnarray*}
0 &=& \overline 0,\\
1 &=& \overline 1.
\end{eqnarray*}
All the other propositions remain distinct.
We then obtain a propositional structure
$$L(x)\oplus L(\overline{x}) =
MO_2$$
whose Hasse diagram is of the ``Chinese lantern'' form and is drawn
in Figure \ref{f-hd-mo2}. The corresponding Greechie Diagram is drawn in
Figure \ref{f-gd-mo2}.
Here, the ``$O$'' stands for {\em orthocomplemented,} the term ``$M$''
stands for {\em modular} (cf. page
\pageref{l-modular} below),
and the subscript ``2'' stands for the pasting of
two Boolean subalgebras $2^2$.
The propositional system obtained is {\em not} a
classical Boolean algebra,
since the distributive law is not satisfied. This can be easily seen by
\index{distributivity}
the following evaluation. Assume that the distributive law
is satisfied. Then,\label{l-dist}
\begin{eqnarray*}
p_- \vee  ({\overline p}_- \wedge  {\overline p}_-') &=&
(p_- \vee  {\overline p}_-) \wedge  (p_- \vee  {\overline p}_-'),\\
p_- \vee  0 &=& 1 \wedge  1,\\
p_-  &=& 1.
\end{eqnarray*}
This is incorrect.
A similar calculation yields
\begin{eqnarray*}
p_- \wedge  ({\overline p}_- \vee  {\overline p}_-') &=&
(p_- \wedge  {\overline p}_-) \vee  (p_- \wedge  {\overline p}_-'),\\
p_- \wedge  1 &=& 0 \wedge  0,\\
p_-  &=& 0,
\end{eqnarray*}
which again is incorrect.

Notice that the expressions can be easily evaluated by using the Hasse
diagram
\ref{f-hd-mo2}.
For any $a,b$,
$a\vee b$ is just the least element which is connected
by $a$ and $b$;
$a\wedge b$ is just the highest element connected
to $a$ and $b$. Intermediates which are not connected to both $a$
and $b$ do not count. That is,
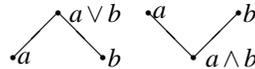
\begin{center}
\unitlength 0.30mm
\linethickness{0.4pt}
\begin{picture}(105.00,20.95)
\put(0.00,0.00){\line(1,1){20.00}}
\put(20.00,20.00){\line(1,-1){20.00}}
\put(60.00,20.00){\line(1,-1){20.00}}
\put(80.00,0.00){\line(1,1){20.00}}
\put(100.00,20.00){\circle*{1.89}}
\put(60.00,20.00){\circle*{1.89}}
\put(20.00,20.00){\circle*{1.89}}
\put(80.00,0.00){\circle*{1.89}}
\put(40.00,0.00){\circle*{1.89}}
\put(0.00,0.00){\circle*{1.89}}
\put(5.00,0.00){\makebox(0,0)[cc]{$a$}}
\put(45.00,0.00){\makebox(0,0)[cc]{$b$}}
\put(65.00,20.00){\makebox(0,0)[cc]{$a$}}
\put(105.00,20.00){\makebox(0,0)[cc]{$b$}}
\put(25.00,20.00){\makebox(0,0)[lc]{$a\vee b$}}
\put(85.33,0.00){\makebox(0,0)[lc]{$a \wedge b$}}
\end{picture}
\end{center}
$a\vee b$ is called a least upper bound of $a$ and $b$.
\index{least upper bound}
$a\wedge b$ is called a greatest lower bound of $a$ and $b$.
\index{greatest lower bound}

$MO_2$ is a specific example of an algebraic structure which is called a
{\em lattice}.
Any two elements of a lattice have a least upper and a greatest lower
bound. Furthermore,
\begin{eqnarray*}
a\rightarrow b
\textrm{ and }
a \rightarrow c, &\textrm{ then }&
a
\rightarrow
(b\wedge c);\\
b
\rightarrow
a
\textrm{ and }
c
\rightarrow
a, &\textrm{ then }&
(b\vee c)
\rightarrow
a.
\end{eqnarray*}

It is an {\em ortholattice} or {\em orthocomplemented lattice}, since
every element has a complement.

It is modular, since for all $a\rightarrow c$, the modular law
\index{modular}
\label{l-modular}
$$
 (a\vee b)\wedge c= a\vee (b \wedge c)
$$
is satisfied. For example,
\begin{eqnarray*}
(p_-\vee p_+)\wedge 1&=& p_-\vee (p_+ \wedge 1),\\
            1\wedge 1&=& p_-\vee p_+,\\
             1&=&1.
\end{eqnarray*}

One can proceed and consider a finite number $n$ of different
directions of spin state measurements, corresponding to $n$ distinct
orientations of a Stern-Gerlach apparatus. The resulting propositional
structure is the horizontal sum $MO_n$ of $n$ classical Boolean algebras
\index{$MO_n$}
$L(x^i)$, where $x^i$ indicates the direction of a spin state
measurement. That is,
$$MO_n=\oplus_{i=1}^n L(x^i).$$
Figure \ref{f-hd-mon} and
Figure \ref{f-gd-mon}
show its
Hasse and
Greechie
diagrams, respectively.
Of course, it should be emphasized that only a {\em single} $L(x^i)$ can
actually be operationalized. According to quantum mechanics, all the
other
$n-1$ unchecked quasi-classical worlds remain in permanent oblivion.
\begin{figure}[htd]
\begin{center}
\unitlength 0.60mm
\linethickness{0.4pt}
\begin{picture}(171.67,126.67)
\put(85.00,0.00){\line(1,4){10.00}}
\put(95.00,40.00){\line(-1,4){9.92}}
\put(85.08,79.67){\line(3,-4){29.75}}
\put(114.83,40.00){\line(-3,-4){30.00}}
\put(84.83,0.00){\line(5,4){50.17}}
\put(135.00,40.13){\line(-5,4){50.00}}
\put(85.00,0.00){\line(5,3){67.00}}
\put(152.00,40.20){\line(-5,3){67.00}}
\put(85.00,80.00){\circle*{3.33}}
\put(95.00,40.00){\circle*{3.33}}
\put(115.00,40.00){\circle*{3.33}}
\put(135.00,40.00){\circle*{3.33}}
\put(152.33,40.00){\circle*{3.33}}
\put(85.00,-0.33){\circle*{3.33}}
\put(93.33,80.00){\makebox(0,0)[lc]{$1$}}
\put(93.33,0.00){\makebox(0,0)[lc]{$0$}}
\put(17.67,35.00){\makebox(0,0)[cc]{$p_-^1$}}
\put(34.67,35.00){\makebox(0,0)[cc]{$p_+^1$}}
\put(55.00,35.00){\makebox(0,0)[cc]{$p_-^2$}}
\put(74.00,35.00){\makebox(0,0)[cc]{$p_+^2$}}
\put(97.00,35.00){\makebox(0,0)[cc]{$p_-^3$}}
\put(115.67,35.00){\makebox(0,0)[cc]{$p_+^3$}}
\put(135.00,35.00){\makebox(0,0)[cc]{$p_-^n$}}
\put(152.67,35.00){\makebox(0,0)[cc]{$p_+^n$}}
\put(14.67,95.00){\circle*{3.33}}
\put(-0.33,110.00){\circle*{3.33}}
\put(29.67,110.00){\circle*{3.33}}
\put(14.67,125.00){\circle*{3.33}}
\put(-0.33,110.00){\line(1,1){15.00}}
\put(14.67,125.00){\line(1,-1){15.00}}
\put(29.67,110.00){\line(-1,-1){15.00}}
\put(14.67,95.00){\line(-1,1){15.00}}
\put(21.00,95.00){\makebox(0,0)[cc]{$0^1 $}}
\put(21.00,125.00){\makebox(0,0)[cc]{$1^1$}}
\put(-0.33,104.67){\makebox(0,0)[cc]{$p_-^1$}}
\put(29.67,105.00){\makebox(0,0)[cc]{$p_+^1$}}
\put(35.50,110.00){\makebox(0,0)[cc]{$\oplus$}}
\put(55.92,95.00){\circle*{3.33}}
\put(98.83,95.00){\circle*{3.33}}
\put(155.00,95.00){\circle*{3.33}}
\put(40.92,110.00){\circle*{3.33}}
\put(83.83,110.00){\circle*{3.33}}
\put(140.00,110.00){\circle*{3.33}}
\put(70.92,110.00){\circle*{3.33}}
\put(113.83,110.00){\circle*{3.33}}
\put(170.00,110.00){\circle*{3.33}}
\put(55.92,125.00){\circle*{3.33}}
\put(98.83,125.00){\circle*{3.33}}
\put(155.00,125.00){\circle*{3.33}}
\put(40.92,110.00){\line(1,1){15.00}}
\put(83.83,110.00){\line(1,1){15.00}}
\put(140.00,110.00){\line(1,1){15.00}}
\put(55.92,125.00){\line(1,-1){15.00}}
\put(98.83,125.00){\line(1,-1){15.00}}
\put(155.00,125.00){\line(1,-1){15.00}}
\put(70.92,110.00){\line(-1,-1){15.00}}
\put(113.83,110.00){\line(-1,-1){15.00}}
\put(170.00,110.00){\line(-1,-1){15.00}}
\put(55.92,95.00){\line(-1,1){15.00}}
\put(98.83,95.00){\line(-1,1){15.00}}
\put(155.00,95.00){\line(-1,1){15.00}}
\put(62.25,95.00){\makebox(0,0)[cc]{$0^2 $}}
\put(105.16,95.00){\makebox(0,0)[cc]{$0^3$}}
\put(161.33,95.00){\makebox(0,0)[cc]{$0^n$}}
\put(62.25,125.00){\makebox(0,0)[cc]{$1^2$}}
\put(105.16,125.00){\makebox(0,0)[cc]{$1^3$}}
\put(161.33,125.00){\makebox(0,0)[cc]{$1^n$}}
\put(40.92,104.67){\makebox(0,0)[cc]{$p_-^2$}}
\put(83.83,104.67){\makebox(0,0)[cc]{$p_-^3$}}
\put(140.00,104.67){\makebox(0,0)[cc]{$p_-^n$}}
\put(70.92,105.00){\makebox(0,0)[cc]{$p_+^2$}}
\put(113.83,105.00){\makebox(0,0)[cc]{$p_+^3$}}
\put(170.00,105.00){\makebox(0,0)[cc]{$p_+^n$}}
\put(77.58,110.00){\makebox(0,0)[cc]{$\oplus$}}
\put(119.67,110.00){\makebox(0,0)[cc]{$\oplus$}}
\put(134.67,110.00){\makebox(0,0)[cc]{$\oplus$}}
\put(121.33,40.00){\circle*{0.67}}
\put(123.00,40.00){\circle*{0.67}}
\put(124.67,40.00){\circle*{0.67}}
\put(125.50,110.00){\circle*{0.67}}
\put(127.17,110.00){\circle*{0.67}}
\put(128.84,110.00){\circle*{0.67}}
\put(5.00,40.00){\makebox(0,0)[cc]{$=$}}
\put(85.00,0.00){\line(-1,4){10.00}}
\put(75.00,40.00){\line(1,4){9.92}}
\put(84.92,79.67){\line(-3,-4){29.75}}
\put(55.17,40.00){\line(3,-4){30.00}}
\put(85.17,0.00){\line(-5,4){50.17}}
\put(35.00,40.13){\line(5,4){50.00}}
\put(85.00,0.00){\line(-5,3){67.00}}
\put(18.00,40.20){\line(5,3){67.00}}
\put(75.00,40.00){\circle*{3.33}}
\put(55.00,40.00){\circle*{3.33}}
\put(35.00,40.00){\circle*{3.33}}
\put(17.67,40.00){\circle*{3.33}}
\end{picture}
\end{center}
\caption{\label{f-hd-mon}
Hasse
diagram of $n$ spin one-half state propositional systems $L(x^i),
i=1,\cdots
,n$
which are noncomeasurable. The superscript $i$ represents the $i$th
measurement direction.
}
\end{figure}
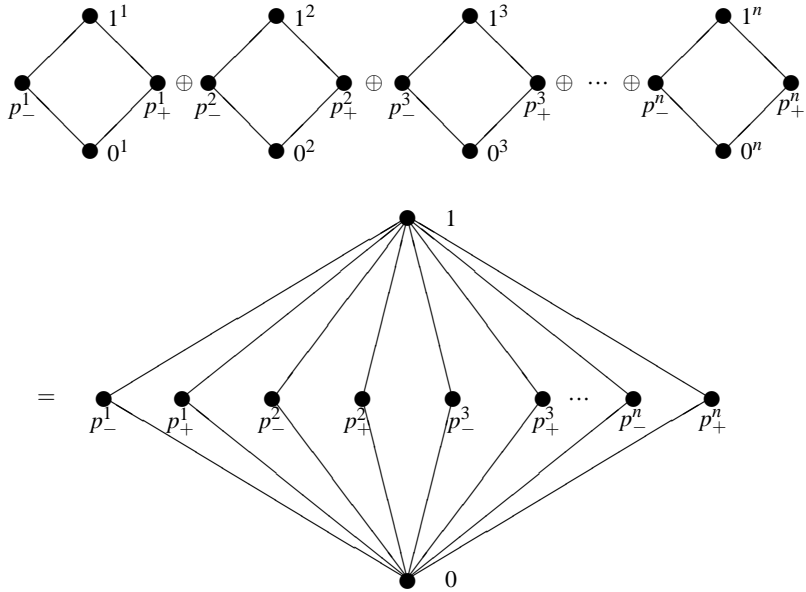
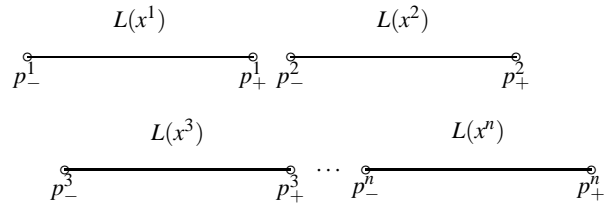
\begin{figure}[htd]
\begin{center}
\unitlength 0.50mm
\linethickness{0.4pt}
\begin{picture}(151.06,45.00)
\put(0.00,35.00){\circle{2.11}}
\put(60.00,35.00){\circle{2.11}}
\put(0.00,30.00){\makebox(0,0)[cc]{$p_-^1$}}
\put(60.00,30.00){\makebox(0,0)[cc]{$p_+^1$}}
\put(0.00,35.00){\line(1,0){60.00}}
\put(70.00,35.00){\circle{2.11}}
\put(130.00,35.00){\circle{2.11}}
\put(70.00,30.00){\makebox(0,0)[cc]{$p_-^2$}}
\put(130.00,30.00){\makebox(0,0)[cc]{$p_+^2$}}
\put(70.00,35.00){\line(1,0){60.00}}
\put(30.00,45.00){\makebox(0,0)[cc]{$L(x^1)$}}
\put(100.00,45.00){\makebox(0,0)[cc]{$L(x^2)$}}
\put(10.00,5.00){\circle{2.11}}
\put(70.00,5.00){\circle{2.11}}
\put(10.00,0.00){\makebox(0,0)[cc]{$p_-^3$}}
\put(70.00,0.00){\makebox(0,0)[cc]{$p_+^3$}}
\put(10.00,5.00){\line(1,0){60.00}}
\put(90.00,5.00){\circle{2.11}}
\put(150.00,5.00){\circle{2.11}}
\put(90.00,0.00){\makebox(0,0)[cc]{$p_-^n$}}
\put(150.00,0.00){\makebox(0,0)[cc]{$p_+^n$}}
\put(90.00,5.00){\line(1,0){60.00}}
\put(40.00,15.00){\makebox(0,0)[cc]{$L(x^3)$}}
\put(120.00,15.00){\makebox(0,0)[cc]{$L(x^n)$}}
\put(80.00,4.67){\makebox(0,0)[cc]{$\cdots$}}
\end{picture}
\end{center}
\caption{\label{f-gd-mon}
Greechie
diagram of $n$ spin one-half state propositional systems $L(x^i),
i=1,\cdots
,n$
which are noncomeasurable. The superscript $i$ represents the $i$th
measurement direction.
}
\end{figure}

\section{Finite subalgebras of $n$--dimensional Hilbert logics}
\label{lx-xxslfdhl2}

The finite subalgebras of two--dimensional Hilbert space are $MO_n,
n\in {\Bbb N}$. This can be visualized easily, since given a vector $v$
associated with a proposition $p_v$, there exists only a {\em single}
orthogonal vector $v'$, corresponding to
the proposition $p_v'$, which is the negation of
the proposition $p_v$. Conversely, the negation of $p_v'$ can be
uniquely identified with the vector $v$.

This is not the case in three- and higher-dimensional spaces, where the
complement of a vector is a plane (or a higher-dimensional
subspace), and is therefore no unique vector.

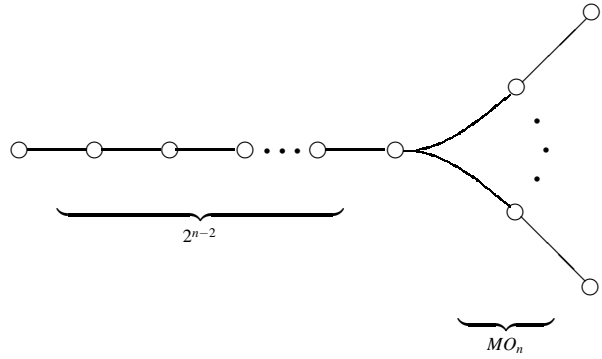
\begin{figure}
\begin{center}
\unitlength 1.00mm
\linethickness{0.4pt}
\begin{picture}(77.13,44.45)
\put(0.00,25.00){\circle{2.11}}
\put(10.00,25.00){\circle{2.11}}
\put(20.00,25.00){\circle{2.11}}
\put(30.00,25.00){\circle{2.11}}
\put(39.66,25.00){\circle{2.11}}
\put(50.00,25.00){\circle{2.11}}
\put(65.90,16.80){\circle{2.11}}
\put(75.90,6.80){\circle{2.11}}
\put(66.07,33.39){\circle{2.11}}
\put(76.07,43.39){\circle{2.11}}
\put(1.10,25.08){\line(1,0){7.83}}
\put(10.97,25.08){\line(1,0){7.83}}
\put(20.84,25.08){\line(1,0){7.83}}
\put(40.74,25.08){\line(1,0){7.83}}
\put(66.89,34.02){\line(1,1){8.51}}
\put(66.72,16.16){\line(1,-1){8.51}}
\put(32.92,24.91){\circle*{0.68}}
\put(34.96,24.91){\circle*{0.68}}
\put(37.00,24.91){\circle*{0.68}}
\put(70.01,25.08){\circle*{0.68}}
\put(68.85,28.85){\circle*{0.68}}
\put(68.85,21.13){\circle*{0.68}}
\put(24.00,15.04){\makebox(0,0)[cc]
{$\underbrace{\qquad \qquad \qquad \qquad \qquad \qquad}_{2^{n-2}}$}}
\put(64.67,0.37){\makebox(0,0)[cc]{$\underbrace{\qquad \qquad}_{MO_n}$}}
\put(51.08,24.92){\line(1,0){1.67}}
\multiput(52.75,25.00)(0.45,0.11){4}{\line(1,0){0.45}}
\multiput(54.56,25.44)(0.28,0.11){7}{\line(1,0){0.28}}
\multiput(56.51,26.23)(0.21,0.11){10}{\line(1,0){0.21}}
\multiput(58.60,27.37)(0.17,0.12){13}{\line(1,0){0.17}}
\multiput(60.82,28.87)(0.15,0.12){30}{\line(1,0){0.15}}
\put(51.08,24.92){\line(1,0){1.67}}
\multiput(52.75,24.84)(0.45,-0.11){4}{\line(1,0){0.45}}
\multiput(54.56,24.40)(0.28,-0.11){7}{\line(1,0){0.28}}
\multiput(56.51,23.61)(0.21,-0.11){10}{\line(1,0){0.21}}
\multiput(58.60,22.47)(0.17,-0.11){13}{\line(1,0){0.17}}
\multiput(60.82,20.98)(0.15,-0.12){30}{\line(1,0){0.15}}
\end{picture}
\end{center}
\caption{\label{f-2n-1xmo2}
Finite subalgebra of $n$--dimensional Hilbert logic
$2^{n-2}\times MO_m$, $1<m\in {\Bbb N}$, $n\ge 3$.}
\end{figure}

The previous results can be generalized to $n$--dimensional Hilbert
spaces. Take, as an incomplete example,
the product of a Boolean algebra of
dimension
$n-2$ and a modular lattice of the Chinese lantern type
$MO_m$ \cite{kalmbach-priv};
e.g.,
\begin{equation}
B\times MO_n=2^{n-2}\times MO_m,\quad 1< m\in {\Bbb N}, \quad n\ge
3.
\label{l-nncsu}
\end{equation}
It is depicted in Figure \ref{f-2n-1xmo2}.

In particular, for $n=3$,
$2^1\times MO_2=L_{12}$.
 In general, $L_{2(2m+2)}=L_{4m+4}=2^1\times MO_m$,
and we are recovering the three-dimensional case discussed before.
\index{$L_{12}$}

The logic $2^{n-2}\times MO_m$ has a
separating set of two-valued states. Therefore, it can be realized by
automaton logics \cite{svozil-93}.

The above class
$2^{n-2}\times MO_m$, $1<m\in {\Bbb N}$ does not coincide with
the class of all modular lattices
corresponding to finite subalgebras of $n$--dimensional Hilbert logics
for $n>3$. Consider, for instance,
four
-dimensional real Hilbert space
${\Bbb R}^4$.
The product $MO_2\times MO_2$ is a subalgebra of the corresponding
Hibert logic but is not a logic represented by Equation
(\ref{l-nncsu}).
This can be demonstrated by identifying the following
eight
one--dimensional subspaces\footnote{
$\sp$ denotes the linear span.}
\begin{eqnarray*}
&a=\sp (1,0,0,0),\quad
a'=\sp (0,1,0,0),\quad
b=\sp (1,1,0,0),\quad
b'=\sp (1,-1,0,0),&\\
&c=\sp (0,0,1,0),\quad
c'=\sp (0,0,0,1),\quad
d=\sp (0,0,1,1),\quad
d'=\sp (0,0,1,-1)&
\end{eqnarray*}
 with the atoms
of the two factors $MO_2$ (four
atoms per factor) \cite{harding-priv}.
Note that $\{a,a',c,c'\}$ and $\{b,b',d,d'\}$ are two
orthogonal tripods in ${\Bbb R}^4$.

This result could be generalized to the product $MO_i \times MO_j$ by
augmenting the above vectors with additional vectors
$(\cos \phi_l,\sin \phi_l,0,0)$,
$(\sin \phi_l,-\cos \phi_l,0,0)$,
$(0,0,\cos \phi_k,\sin \phi_k)$,
$(0,0,\sin \phi_k,-\cos \phi_k)$, such that all angles $\phi_{l,k}$ are
mutually different,
$l,k\in {\Bbb N}$, and
$1<l \le i$,
$1<l \le j$.

Furthermore,
the above considerations could be extended
for even-dimensional vector spaces by the proper multiplication of
additional $MO_2$ ($MO_m$) factors. For instance, for six--dimensional
Hilbert logic, we may consider three factors $MO_2$ corresponding to
\begin{eqnarray*}
& (1,0,0,0,0,0),
(0,1,0,0,0,0),
(1,1,0,0,0,0),
 (1,-1,0,0,0,0),&\\
& (0,0,1,0,0,0),
(0,0,0,1,0,0),
(0,0,1,1,0,0),
(0,0,1,-1,0,0),&\\
& (0,0,0,0,1,0),
(0,0,0,0,0,1),
(0,0,0,0,1,1),
(0,0,0,0,1,-1),
\end{eqnarray*}
each one of the three rows describing the atoms of one of the factors.
Indeed,
if $L(V)$ is the
(ortho) lattice of subspaces of $V$, then
the map $f:L(V) \times L(W) \rightarrow L(V x W)$ defined by
$f((S,T)) = S \times T$ is an (ortho) lattice embedding. This is what
makes the $MO_2 \times MO_2$ example work \cite{harding-priv}.

Let us denote by ${\frak C} ({\Bbb R}^n)$ the orthomodular lattice of all
subspaces of
${\Bbb R}^n$. This
lattice is  modular. Furthermore, any sublattice  of  ${\frak C} ({\Bbb
R}^n)$
is modular. As has been pointed out by Chevalier \cite{cheval-priv},
\begin{itemize}

         \item  any finite orthomodular lattice is isomorphic to a
direct product $B
\times
        \displaystyle \prod_{i\in I}L_i$  where $B$ is a Boolean algebra and
         the  $L_i$ are simple (not isomorphic to a product)
orthomodular lattices.
        \item   The simple finite modular ortholattices are
the Boolean algebra
$2^1=\{0,1\}$
          and the $MO_n$, $n\geq 2$. Thus any finite modular
ortholattice is isomorphic to
        $2^n \times MO_{n_1}\times \dots\times MO_{n_k}$.
        \item  If $a_1,\dots, a_n$ are pairwise
        orthogonal elements of an orthomodular lattice
        $L$, such that   $a_1 \vee \dots \vee a_k =1$,
        then the direct product
         $\prod [0,a_i]$ is isomorphic to a sub-orthomodular lattice
        of $L$. If the $a_i$ are not central elements then this sub-orthomodular lattice is not
        equal to $L$.
         \item If $n= n_1 +\dots+n_k$, $n_i>0$, then there exist in
${\Bbb R}^n$
         pairwise orthogonal subspaces $M_1,\dots,M_k$ such that ${\Bbb
R}^n = M_1 \vee
          \dots \vee M_k$ and $dim M_{n_i} = n_i$. Thus ${\frak C}
({\Bbb R}^{n_1})\times
         \dots \times  {\frak C} ({\Bbb R}^{n_k})$ is a sub-orthomodular lattice of ${\frak C} ({\Bbb R}^n)$.
        \item  Any sub-orthomodular lattice of a  sub-orthomodular lattice is a sub-orthomodular lattice.
        \item  $2^1=\{0 ,1\}$ is a sub-orthomodular lattice
        of ${\frak C} ({\Bbb R}^n)$ if $n>0$ and $2^p $ is a
sub-orthomodular lattice of ${\frak C}
        ({\Bbb R}^n)$ iff $p\leq n$.
        \item  Any $MO_p$ is a sub-orthomodular lattice of ${\frak C} ({\Bbb R}^{2n})$ but is not a
         sub-orthomodular lattice of ${\frak C} ({\Bbb R}^{2n+1})$. The
reason is: In $MO_p$, 1 is a commutator. 1 is
        not a commutator in ${\frak C} ({\Bbb R}^{2n+1})$
\cite{cheval-or}.
 For
        the same reason, a product of $MO_p$, without a Boolean factor,  is not
        a sub-orthomodular lattice of ${\frak C} ({\Bbb R}^{2n+1})$.
\end{itemize}

Let $n>0$ be an integer.
          The finite sub-orthomodular posets of ${\frak C} ({\Bbb
R}^n)$ are the
        $$2^q \times MO_{n_1} \times\dots \times MO_{n_k}\quad
\textrm{with}\quad
q +2k\leq
        n.$$ If $n$ is odd then $q$ must be different from 0 (the
product must
        contain a Boolean factor) \cite{cheval-priv}.\footnote{
Notice that $MO_2\times MO_2 $ is, for example,  not a sub-orthomodular
poset of
${\frak C} ({\Bbb R}^5)$.}

Members of this class also
have a separating set of two-valued states.

        Any finite modular orthomodular lattice is isomorphic to a
sub-orthomodular lattice of some
${\frak C}
({\Bbb R}^n)$.


\section{Probabilities}

\subsection{Framework}

Let us first programmatically state the framework in which we wish to operate.
In our view, probabilities are quantitative expectations of experimental frequencies
of certain observed events.
The most fundamental yes-no event or outcome in quantum mechanics is a detector click:
there either is such a click, or there is none.
These events are all that ``we have;'' there is no other empirical physical
entity or property, such as human intuition, which can be justifiable called
``physical.''
There is no such thing as certainty in physics (and life in general ;-);
thus all physical experience inevitably is stochastic.

Induction is ``bottom-up.'' It attempts to reconstruct certain postulated (quantum)
structures from such elementary events.
The induction problem, in particular effective algorithmic ways and methods
to derive certain outcomes or events from other (causally ``previous'')
events or outcomes via some kind of ``narratives'' such as physical theories,
still remains unsolved.
Indeed, in view of powerful formal incompleteness theorems, such as the halting problem,
the busy beaver function,
or the recursive unsolvability of the
rule inference problem, the induction problem is provable recursively unsolvable for
physical systems which can be reduced to, or at least contain, universal Turing machines.
The physical universe as we know it, appears to be of that kind
(cf. Refs.~\cite{svozil-93,svozil-unev}).

Deduction is ``top-down.'' It postulates certain entities such as physical theories.
Those theories may just have been provided by an oracle,
they may be guesswork or just random pieces of data crap in a computer memory.
Deduction then derives empirical consequences from those theories.

In both cases, probabilities are the only interface between physical theories
and the richness of physical experiences which ultimately seem to consist of
elementary yes-no events or outcomes.

\subsection{Classical probabilities}

Classical probabilities are probabilities about events or outcomes of classical systems.
Classical systems are Boolean; e.g., distributive, by definition.

In what follows, the terms valuation, two-valued (probability) measure, as well as
dispersionless state and classical truth assignment will be used as synonyms.
Suppose a classical Boolean algebra has $n$ elements.
Then there exist $2^n$ such truth assignments.
All classical probabilities can be represented as a convex sums
of all the possible truth assignments; this is a necessary and sufficient condition.

As a consequence, there exist certain constraints on classical probabilities; constraints which
already were considered by Boole \cite{Boole,Boole-62} over 150 years ago.
In order to establish bounds on quantum probabilities,
Pitowsky (among others; see, e.g., Froissart \cite{froissart-81} and
Tsirelson  \cite{cirelson:80,cirelson})
has given a geometrical interpretation of the bounds of classical probabilities
in terms of correlation polytopes \cite{pitowsky-86,pitowsky,pitowsky-89a,Pit-91,Pit-94}.

Consider an arbitrary number of classical events or outcomes $a_1, a_2,\ldots , a_n$.
Take some (or all of) their probabilities
and some (or all of) the joint probabilities
$p_1, p_2,\ldots , p_n, p_{12},\ldots $
and identify them with the components of
a vector  $p=(p_1, p_2,\ldots , p_n, p_{12},\ldots )$
formed in Euclidean space.
Suppose that the  events or outcomes
$a_1, a_2,\ldots , a_n$
are independent.
Then the
probabilities $p_i$, $i=1,\ldots ,n$ may acquire
the extreme cases $0,1$ independently.
Consider all vectors spanned by those ``extremal'' components.
The combined values of $p_1, p_2,\ldots , p_n$ of the extreme cases $p_i=0,1$,
together with the joined probabilities $p_{ij} =p_i p_j$
can also be interpreted as rows of a truth table; with $0,1$ corresponding to
``{\it false}'' and
``{\it true,}'' respectively.
Moreover, any such entry corresponds to a {\em two-valued measure}
(also called {\em valuation, 0-1-measure} or {\em dispersionless measure}).

In geometrical terms,
any classical probability distribution is representable by some convex sum over
all two-valued measures characterized by the row entries of the truth tables.
That is, it corresponds to
some point on the face of the classical correlation polytope $C={\rm conv} (K)$
which is defined by the set of all points whose
convex sum extends over all vectors associated with row entries in the truth table $K$.
More precisely,
consider the convex hull
${\rm conv} (K)=\left\{ \sum_{i=1}^{2^n} \lambda_i{\bf x}_i
  \; \left|  \;
\lambda_i\ge 0,\; \sum_{i=1}^{2^n}\lambda_i =1
\right.
\right\} $
of the set
$$K
=\{{\bf x}_1,{\bf x}_2,\ldots ,{\bf x}_{2^n}\}
= \left\{
\left.
\large(t_1, t_2,\ldots , t_n, t_xt_y,\ldots \large)
\; \right| \;
t_i \in \{0,1\},\; i=1,\ldots ,n
\right\}.$$
Here, the terms $t_xt_y,\ldots$ stand for arbitrary products associated with
the joint propositions which are considered. Exactly what terms  are
considered depends on the particular physical configuration.

By the Minkoswki-Weyl representation theorem \cite[p.29]{ziegler},
every convex polytope has a dual (equivalent) description:
(i)
either as the convex hull of its extreme points; i.e., vertices;
(ii)
or as the intersection of a finite number of half-spaces,
each one given by a linear inequality.
The linear inequalities,
which are obtained
from the set $K$ of vertices
by solving the so called {\em hull problem}
coincide with Boole's ``conditions of possible experience.''

For particular physical setups,
the inequalities can be identified with Bell-type inequalities which have to be satisfied by
all classical probability distributions.
These conditions are demarcation criteria; i.e.,
they are complete and maximal in the sense that no other system of inequality exist
which characterizes the correlation polytopes completely and exhaustively.
(That is, the bounds on probabilities cannot be enlarged and improved).
Generalizations to the joint distributions of more than two particles are straightforward.
Correlation polytopes have provided a systematic, constructive way of finding
the entire set of Bell-type inequalities associated with any particular physical configuration
\cite{2000-poly,2001-cddif}, although from a computational complexity point of view
\cite{garey}, the problem remains intractable \cite{Pit-91}.

\subsection{Probabilities for nonclassical propositional structures}

Presently nobody knows how to systematically implement probabilities on nonboolean, nonclassical propositional structures.
Some attempts have been made for logics which have ``enough,''
i.e., a full, separating set of two-valued states, which makes possible to allow a faithful embedding into Boolean algebras \cite{svozil-ql}.
Such structures emerge, for instance, in finite automata \cite{svozil-93,schaller-96,dvur-pul-svo,cal-sv-yu,svozil-ql}, or in generalized urn models  \cite{wright:pent,wright}.

The quantum probabilities can be either be postulated by the Born rule (see below), or by Gleason's theorem, which
requires the quasi-classicality of probabilities on all the possible classical mini-universes
characterized ba the maximal set of mutually commuting operators (or simply by a single maximal operators).

\subsection{Quantum probabilities}

Quantum logic suggests that the classical Boolean propositional
structure of events should be replaced by the Hilbert
lattice $\ll$ of subspaces
of a Hilbert space $\hh$. (Alternatively, we may use
the set of all projection operators $\ppp$.)
Thus we should be able to define a probability measure on subspaces of a
Hilbert space  as a normed function
$P$ which assigns to every subspace a nonnegative real number such that
if $\{\mm_{p_i} \}$ is any countable set of mutually orthogonal
subspaces   (corresponding to comeasurable propositions $p_i$) having
closed linear span $\mm_{\vee_i p_i}=\oplus_i \mm_{p_i}$, then
\begin{equation}
P(\mm_{\vee_i p_i})=\sum_i P(\mm_{p_i}). \label{l-ka1}
\end{equation}
Furthermore, the tautology corresponding to the entire Hilbert space
should have probability one. That is,
\begin{equation}
P({\frak H})=1. \label{l-ka2}
\end{equation}
Instead of the subspaces, we could have considered the corresponding
projection operators.

A measure of the above type can be obtained by selecting
an arbitrary normalized vector $ y\in \hh$ and by identifying
$P_{y}(\mm_x)$ with the square of the absolute value of the scalar
product of the orthogonal projection of $y$ onto $\mm_x$ spanned by the
unit vector $x$,
$$P_{ y}(\mm_x) =\vert (x,y)\vert^2 .$$
More generally, any linear combination
$\sum_i P_{ x_i}$
of such
measures $P_{ x_i}$ is again such a measure.
In what follows we shall refer to states corresponding to
one--dimensional subspaces or projections
as {\em pure states}.
\index{pure state}

Indeed, a celebrated theorem by Gleason \cite{Gleason} states
that
in
a separable Hilbert space of dimension at least three,
{\em every} probability measure on the projections
satisfying (\ref{l-ka1})  and (\ref{l-ka2})
can be written in the form
\begin{equation}
\label{l-gleason}
P_\rho (E_\mm )=\textrm{trace}(\rho E_\mm).
\end{equation}
$E_\mm$ denotes the orthogonal projection on $\mm$ and
$\rho $ is a unique positive (semi-definite) self-adjoint {\em density
operator} of the trace class;
\index{density operator}
i.e.,
$(\rho x,x)=(x,\rho x)\ge 0$ for all $x\in \hh$, and
$\textrm{trace}(\rho )=1$.

Then the {\em expectation value}
\index{expectation value}
of an observable corresponding to a self-adjoint operator $A$ with
eigenvalues $\lambda_i$ is
(in $n$--dimensional Hilbert space) given by
\begin{equation}
\label{l-gleason1}
\langle A\rangle =
\sum_{i=1}^n \lambda_i P(E_i) =
\sum_{i=1}^n \lambda_i \textrm{trace}(\rho E_i)=
\textrm{trace}(\rho A).
\end{equation}
The relationship between
(\ref{l-gleason})       and
(\ref{l-gleason1})
is due to the spectral decomposition of $A$ .

Gleason's theorem can be seen as a {\it substitute} for the probability
axiom of quantum mechanics by deriving it from some ``fundamental''
assumptions and
``reasonable'' requirements.
One such requirement is that, if $E_p$ and
$E_q$ are orthogonal projectors representing comeasurable, independent
propositions $p$ and $q$, then their join $p\vee q$ (corresponding to
$E_p+E_q$) has probability
$P(p\vee q)=P(p)+P(q)$ (corresponding to
$P(E_p+E_q)=P(E_p)+P(E_q)$).

\section{Contextuality}

In the late 50's, Ernst Specker
was considering
the question of whether it might be possible to consistently define
elements of physical reality
 {\em ``globally''} which can merely be measured {\em
``locally''} \cite{specker-60}.
Specker mentions the scholastic
speculation of the
so-called ``infuturabilities''; that is, the question of {\em whether
\index{infuturabilities}
the
omniscience (comprehensive knowledge) of God extends to events which
would have occurred if something  had happened which did not
happen}
(cf.
\cite[p. 243]{specker-60} and
\cite[p. 179]{specker-ges}).
Today, the scholastic term
``infuturability'' would be called ``counterfactual.''

\begin{figure}
\begin{center}
\unitlength 0.80mm
\linethickness{0.4pt}
\begin{picture}(107.62,83.81)
\put(15.48,83.81){\line(1,-1){10.00}}
\put(25.48,73.81){\line(-1,-3){5.00}}
\put(20.48,58.81){\line(-3,-2){15.00}}
\put(5.48,48.81){\line(-1,3){5.00}}
\put(0.48,63.81){\line(1,3){5.00}}
\put(5.48,78.81){\line(2,1){10.00}}
\put(15.48,73.81){\line(-1,-2){5.00}}
\put(10.48,63.81){\line(1,-1){5.00}}
\put(10.48,63.81){\line(-1,0){5.00}}
\put(84.05,15.24){\line(-1,1){10.00}}
\put(74.05,25.24){\line(-3,-1){15.00}}
\put(59.05,20.24){\line(-2,-3){10.00}}
\put(49.05,5.24){\line(3,-1){15.00}}
\put(64.05,0.24){\line(3,1){15.00}}
\put(79.05,5.24){\line(1,2){5.00}}
\put(74.05,15.24){\line(-2,-1){10.00}}
\put(64.05,10.24){\line(-1,1){5.00}}
\put(64.05,10.24){\line(0,-1){5.00}}
\put(50.48,73.81){\line(-1,-2){5.00}}
\put(45.48,63.81){\line(2,-1){20.00}}
\put(65.48,53.81){\line(-1,3){5.00}}
\put(60.48,68.81){\line(3,1){15.00}}
\put(75.48,73.81){\line(-3,2){15.00}}
\put(60.48,83.81){\line(-1,0){10.00}}
\put(50.48,83.81){\line(0,-1){10.00}}
\put(55.48,78.81){\line(0,-1){10.00}}
\put(55.48,68.81){\line(1,-1){5.00}}
\put(55.48,68.81){\line(-1,-1){5.00}}
\put(33.81,14.29){\line(1,2){5.00}}
\put(38.81,24.29){\line(-2,1){20.00}}
\put(18.81,34.29){\line(1,-3){5.00}}
\put(23.81,19.29){\line(-3,-1){15.00}}
\put(8.81,14.29){\line(3,-2){15.00}}
\put(23.81,4.29){\line(1,0){10.00}}
\put(33.81,4.29){\line(0,1){10.00}}
\put(28.81,9.29){\line(0,1){10.00}}
\put(28.81,19.29){\line(-1,1){5.00}}
\put(28.81,19.29){\line(1,1){5.00}}
\put(30.48,53.81){\line(-1,-3){5.00}}
\put(25.48,38.81){\line(5,-1){25.00}}
\put(50.48,33.81){\line(1,1){10.00}}
\put(60.48,43.81){\line(-4,3){20.00}}
\put(40.48,58.81){\line(-2,-1){10.00}}
\put(35.48,48.81){\line(2,-1){10.00}}
\put(45.48,43.81){\line(1,1){5.00}}
\put(45.48,43.81){\line(0,-1){5.00}}
\put(10.71,63.81){\line(1,2){5.00}}
\put(10.24,63.81){\line(1,2){5.00}}
\put(45.48,43.57){\line(-2,1){10.00}}
\put(45.48,44.05){\line(-2,1){10.00}}
\put(55.00,68.57){\line(-1,-1){4.76}}
\put(50.71,63.81){\line(1,1){4.76}}
\put(64.20,10.04){\line(2,1){10.24}}
\put(64.05,10.48){\line(2,1){10.00}}
\put(28.57,19.52){\line(0,-1){10.71}}
\put(29.05,8.81){\line(0,1){10.71}}
\put(100.00,50.00){\line(0,-1){10.00}}
\put(100.00,40.00){\line(5,-2){7.62}}
\put(100.00,40.02){\line(-5,-3){5.95}}
\put(98.32,33.33){\line(1,4){1.67}}
\put(99.98,40.00){\line(-2,1){5.71}}
\put(97.38,45.24){\line(1,-2){2.62}}
\put(100.00,40.00){\line(2,1){5.24}}
\put(75.00,40.00){\vector(1,0){10.00}}
\put(80.00,45.00){\makebox(0,0)[cc]{?}}
\put(37.14,72.86){\makebox(0,0)[cc]{$\cdots$}}
\put(50.00,23.81){\makebox(0,0)[cc]{$\cdots$}}
\put(74.05,61.90){\makebox(0,0)[cc]{$\cdots$}}
\put(6.67,30.71){\makebox(0,0)[cc]{$\cdots$}}
\put(102.86,35.24){\makebox(0,0)[cc]{$\cdots$}}
\put(103.33,46.19){\makebox(0,0)[cc]{$\ddots$}}
\end{picture}
\end{center}
\caption{\label{f-jigsaw}
The quantum jigsaw puzzle in three dimensions: is it possible to
consistently arrange
undenumerably many pieces of counterfactual observables, only one of
which is actually measurable? Every tripod has a red leg (thick
line) and two green legs.}
\end{figure}
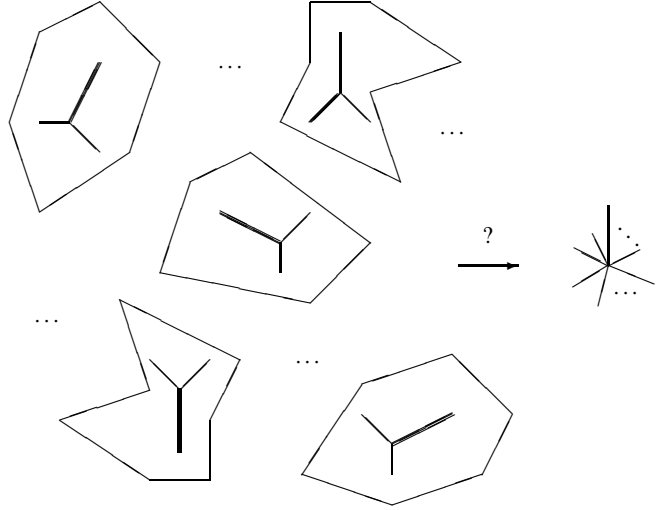

Let us be more specific.
Here, the meaning of the terms local and global will be
understood as follows.
In quantum mechanics, every single orthonormal basis of a Hilbert space
corresponds to locally comeasurable elements of physical reality.
The (undenumerable) class of all orthonormal basis of a Hilbert
space
corresponds to a global description of the conceivable
observables --- Schr\"odinger's catalogue of expectation values
\cite{schrodinger}.
It is quite reasonable to ask whether one could (re)construct the
global description from its
single, local, parts, whether the pieces could be used to
consistently
define the whole. A metaphor of this motive is the quantum jigsaw puzzle
depicted in Figure \ref{f-jigsaw}.
In this jigsaw puzzle, all legs should be translated to the origin.
Every single piece of the jigsaw puzzle consists of
mutually orthogonal rays. It has exactly
one
``privileged'' leg, which
is singled out by coloring it differently from the other (mutual)
orthogonal legs (or, alternatively, assigning to it the probability
measure one, corresponding to certainty).
The pieces should be arranged such that one and the same leg occurring
in two or more pieces should have the same color (probability measure)
for every piece.

As it turns out, for Hilbert spaces of dimension greater than two, the
jigsaw puzzle is unsolvable. That is, every attempt to arrange the
pieces consistently into a whole is doomed to fail. One characteristic
of this failure is that {\em legs (corresponding
to elementary propositions) appear differently
colored, depending on the particular tripod they are in!} More
explicitly: there may exist two tripods (embedded in a
larger tripod set) with one common leg, such that this leg appears red
in one tripod and green in the other one.
Since every tripod is associated with a system of
mutually compatible observables,
this could be interpreted as an indication that the truth or falsity of
a proposition (and hence the element of physical reality)
associated with
it depends on the {\em context} of measurement
\cite{bell-66,redhead}
\index{context dependence}; i.e., whether it is measured along with
first or second frame of mutually compatible
observables.\footnote{Measurement of propositions corresponding
to a given triad can be reduced to a single ``Ur''-observable per
triad.}
It is in this sense that the nonexistence of two-valued probability
measures is
a formalization of the concept of context dependence or contextuality.
\index{contextuality}

Observe that at this point, the theory takes an
unexpected turn. The whole issue of a ``secret
classical arena beyond quantum mechanics'', more specifically
noncontextual hidden parameters, boils down to a consistent coloring of a
finite number of vectors in three--dimensional space!

One of the most compact and comprehensive versions of the Kochen-Specker
argument in three--dimensional Hilbert space ${\Bbb R}^3$ has been given
by Peres
\cite{peres-91}.
(For other discussions, see Refs.
\cite{stairs83,redhead,jammer-92,brown,peres-91,peres,penrose-ks,clifton-93,mermin-93,svozil-tkadlec}.)
Peres' version uses
a 33-element
set of lines without any two-valued state.
The direction vectors of these lines
arise by all permutations of coordinates from
\begin{equation}
(0,0,1),\;
(0,\pm 1,1),\;
(0,\pm 1,\sqrt{2}),\;
\;\textrm{ and }\; (\pm 1,\pm 1,\sqrt{2}).
\label{p-ksl}
\end{equation}
As will be explicitly enumerated below, these lines can be generated (by
the {\it nor}-operation between nonorthogonal propositions)
by the three lines
\cite{svozil-tkadlec}
$$(1,0,0),\; (1,1,0),\; (\sqrt2,1,1).$$
Note that as three arbitrary but mutually nonorthogonal lines generate
a dense set of lines
(cf. \cite{havlicek}),
it can be expected that any such triple of lines (not just the one
explicitly mentioned) generates a finite set of lines which does not
allow a two-valued probability measure.

The way it is defined, this set of lines is invariant under interchanges
(permutations) of the $x_1,x_2$ and $x_3$ axes, and under a reversal of
the direction of each of these axes.
This symmetry property allows us to assign the probability measure $1$
to some of the rays without loss of generality --- assignment of
probability measure $0$ to these rays would be equivalent to renaming
the axes, or reversing one of the axes.

The Greechie diagram
of the Peres configuration
is given in Figure~\ref{sk33} \cite{svozil-tkadlec}.
For simplicity, 24~points which
belong to exactly one edge are  omitted.
 The coordinates should be read as follows:
$\bar{1}\rightarrow -1$ and $2\rightarrow \sqrt{2}$; e.g.,
1$\bar{1}$2
 denotes
$\textrm{Sp} (1,-1,\sqrt{2})$.
 Concentric circles indicate the (non
 orthogonal) generators mentioned above.
\newsavebox{\vertex}\savebox{\vertex}{
  {\unitlength1mm\begin{picture}(0,0)\put(0,0){\circle{1}}\end{picture}}}
\begin{figure}[p]
\unitlength .45\textwidth
\newsavebox{\subdiagram}\savebox{\subdiagram}{
{\unitlength1mm\begin{picture}(0,0)\put(0,0){\circle{2}}\end{picture}}}
\newcommand{\discbig}{\usebox{\subdiagram}}
\newcommand{\emlin}[4]%
  {\put(#1,#2){\special{em:moveto}}\put(#3,#4){\special{em:lineto}}}
\newcommand{\disc}{\usebox{\vertex}}
\newcommand{\place}[6]%
  {\put(#1,#2){\hspace{#3pt}\raisebox{#4pt}{\makebox(0,0)[#5]{$#6$}}}}
 \newcommand{\point}[2]{\put(#1,#2){\disc}}
\newcommand{\onethird}[3]{\axis#1\side#1#2\side#1#3\cross#2#3}
\newcommand{\axis}[4]%
  {\point#1\point#2\point#3\point#4\emlin#1#3\emlin#3{0}{0}}
\newcommand{\side}[8]%
  {\point#5\point#6\point#7\point#8\emlin#1#5\emlin#5#6\emlin#2#8\emlin#8#7}
\newcommand{\cross}[8]{\emlin#1#7\emlin#7#3\emlin#3#5\emlin#2{0}{0}}
\catcode`\!=\active  \def!{\bar1}
\begin{center}
\begin{picture}(2,2)(-1,-1)
\onethird
{{{ 0.100}{-0.995}}{{ 0.050}{-0.747}}{{ 0.000}{-0.500}}{{ 0.000}{-0.250}}}%
{{{ 0.643}{-0.766}}{{ 0.087}{-0.050}}{{ 0.100}{-0.250}}{{ 0.536}{-0.112}}}%
{{{-0.643}{-0.766}}{{-0.087}{-0.050}}{{-0.100}{-0.250}}{{-0.536}{-0.112}}}%
\onethird
{{{ 0.812}{ 0.584}}{{ 0.622}{ 0.417}}{{ 0.433}{ 0.250}}{{ 0.217}{ 0.125}}}%
{{{ 0.342}{ 0.940}}{{-0.000}{ 0.100}}{{ 0.167}{ 0.212}}{{-0.171}{ 0.520}}}%
{{{ 0.985}{-0.174}}{{ 0.087}{-0.050}}{{ 0.267}{ 0.038}}{{ 0.365}{-0.408}}}%
\onethird
{{{-0.912}{ 0.411}}{{-0.672}{ 0.330}}{{-0.433}{ 0.250}}{{-0.217}{ 0.125}}}%
{{{-0.985}{-0.174}}{{-0.087}{-0.050}}{{-0.267}{ 0.038}}{{-0.365}{-0.408}}}%
{{{-0.342}{ 0.940}}{{-0.000}{ 0.100}}{{-0.167}{ 0.212}}{{ 0.171}{ 0.520}}}%
\put (0,0){\circle{0.2}}
\put ( 0.000, 0.100){\discbig}
\put ( 0.050,-0.747){\discbig}
\put ( 0.217, 0.125){\discbig}
\footnotesize
\place { 0.100}{-0.995}{ 0}{-6}{ t}{2!!}
\place { 0.050}{-0.747}{ 6}{ 0}{l }{211}
\place { 0.000}{-0.500}{ 6}{ 0}{l }{01!}
\place { 0.000}{-0.250}{ 2}{ 3}{lb}{011}
\place { 0.000}{ 0.100}{ 0}{-4}{ t}{100}
\place { 0.100}{-0.250}{ 4}{-4}{lt}{2!1}
\place {-0.100}{-0.250}{-4}{-4}{rt}{21!}
\place { 0.643}{-0.766}{ 6}{ 0}{l }{102}
\place { 0.365}{-0.408}{ 6}{ 0}{l }{20!}
\place { 0.087}{-0.050}{ 4}{ 0}{lb}{010}
\place {-0.643}{-0.766}{-6}{ 0}{r }{120}
\place {-0.365}{-0.408}{-6}{ 0}{r }{2!0}
\place {-0.087}{-0.050}{-4}{ 0}{rb}{001}
\place { 0.812}{ 0.584}{ 6}{ 0}{lb}{!!2}
\place { 0.622}{ 0.417}{ 6}{ 0}{lt}{112}
\place {-0.912}{ 0.411}{-6}{ 0}{rb}{!2!}
\place {-0.672}{ 0.330}{-6}{ 0}{rt}{121}
\place { 0.433}{ 0.250}{ 3}{-3}{lt}{1!0}
\place { 0.217}{ 0.125}{-6}{ 1}{r }{110}
\place {-0.433}{ 0.250}{-3}{-3}{rt}{10!}
\place {-0.217}{ 0.125}{ 6}{ 1}{l }{101}
\place { 0.267}{ 0.038}{ 6}{ 0}{l }{1!2}
\place { 0.167}{ 0.212}{ 0}{ 6}{lb}{!12}
\place {-0.267}{ 0.038}{-6}{ 0}{r }{12!}
\place {-0.167}{ 0.212}{ 0}{ 6}{rb}{!21}
\place { 0.985}{-0.174}{ 4}{-6}{ t}{201}
\place { 0.536}{-0.112}{ 0}{-6}{lt}{!02}
\place {-0.985}{-0.174}{-4}{-6}{ t}{210}
\place {-0.536}{-0.112}{ 0}{-6}{rt}{!20}
\place { 0.342}{ 0.940}{ 0}{ 6}{ b}{021}
\place { 0.171}{ 0.520}{-4}{ 4}{rb}{0!2}
\place {-0.342}{ 0.940}{ 0}{ 6}{ b}{012}
\place {-0.171}{ 0.520}{ 4}{ 4}{lb}{02!}
\end{picture}
\end{center}
 \caption{Greechie diagram of a set of propositions
embeddable in ${\Bbb R}^3$
 without any two-valued probability measure
  \protect\cite[Figure 9]{svozil-tkadlec}.
\label{sk33}}
\end{figure}

Let us prove that there is no two-valued probability measure
\cite{svozil-tkadlec,tkadlec-96}.
Due to the symmetry of the problem, we can  choose a particular
coordinate axis such that, without loss of generality,
$P(100)=1$.
Furthermore, we may assume (case 1) that
$P(21\bar{1}) = 1$.
It immediately follows that $P(001) = P(010) =
P(102) = P(\bar{1}20) = 0$.
A second glance shows that $P(20\bar{1}) = 1$, $P(1\bar{1}2) = P(112) = 0$.

Let us now suppose (case 1a)
 that $P(201) = 1$. Then  we obtain $P(\bar{1}12) = P(\bar{1}\bar{1}2)
= 0$. We are forced to accept $P(110)
= P(1\bar{1}0)  = 1$ --- a contradiction, since $(110)$ and
$(1\bar{1}0)$ are
orthogonal to each other and lie on one edge.

Hence we have to assume (case 1b) that $P(201) = 0$.
This gives immediately
$P(\bar{1}02)=1$ and
$P(211) =0$.  Since $P(01\bar{1})=0$, we obtain $P(2\bar{1}\bar{1})=1$ and thus
$P(120)=0$.
This requires $P(2\bar{1}0)=1$ and therefore $P(12\bar{1})=P(121)=0$.
Observe that $P(210) = 1$, and thus $P(\bar{1}2\bar{1}) = P(\bar{1}21) = 0$.
In the following step, we notice that $P(10\bar{1}) =
P(101) = 1$ --- a contradiction,
since $(101)$ and $(10\bar{1})$ are
orthogonal to each other and lie on one edge.

Thus we are forced to assume (case 2) that
$P(2\bar{1}1) = 1$. There is no third alternative, since $P(011)=0$ due to the
orthogonality with $(100)$. Now we can repeat the argument for case 1 in
its mirrored form.


The above mentioned set of lines
(\ref{p-ksl})
orthogenerates (by the {\it nor}-opera\-tion between orthogonal vectors)
a suborthoposet of~${\Bbb R}^3$ with 116~elements; i.e., with 57~atoms
corresponding to one--dimensional subspaces spanned by the vectors just
mentioned --- the direction
vectors of the remaining lines arise by all permutations of coordinates
from
$(\pm 1,\pm 1,\sqrt{2})$ --- plus  their two--dimensional orthogonal
planes
plus the entire Hilbert space and the null vector
\cite{svozil-tkadlec}.

This suborthoposet of~${\Bbb R}^3$ has a 17-element set of
orthogenerators; i.e; lines with direction vectors $(0,0,1)$,  $(0,1,0)$
and all coordinate permutations from $(0,1,\sqrt2)$, $(1,\pm1,\sqrt2)$.
It has a
3-element set of generators
$$(1,0,0),\; (1,1,0),\; (\sqrt2,1,1).$$
More explicitly,
{
\let\charmi=A\let\charmii=B      
\newcommand{\ex}[1]%
  {\let\char#1\ifx\char\charmi-1\else\ifx\char\charmii-2\else#1\fi\fi}
\newcommand{\gen}[9]{\textrm{Sp}(\ex#1,\ex#2,\ex#3)&=&
  (\textrm{Sp}(\ex#4,\ex#5,\ex#6)\oplus     \textrm{Sp}(\ex#7,\ex#8,\ex#9))' \equiv
\\*&&}
\def\sqrtii{\sqrt2}\catcode`\2=\active\def2{\sqrtii}
\catcode`\+=\active\def+{\, \textit{nor}\,}
\begin {eqnarray*}
\textrm{Sp}(1,0,0) &=& a ,\\
\textrm{Sp}(1,1,0) &=& b ,\\
\textrm{Sp}(2,1,1) &=& c ,\\
\gen 001100110 (a+b),\\
\gen 01A100211 (a+c),\\
\gen 010100001 (a+(a+b)),\\
\gen 01110001A (a+(a+c)),\\
\gen 1A0110001 (b+(a+b)),\\
\gen A20211001 (c+(a+b)),\\
\gen 2AA21101A (c+(a+c)),\\
\gen A02211010 (c+(a+(a+b))),\\
\gen 210001A20 ((a+b)+(c+(a+b))),\\
\gen 1200012AA ((a+b)+(c+(a+c))),\\
\gen 1020102AA ((a+(a+b))+(c+(a+c))),\\
\gen 21A011A20 ((a+(a+c))+(c+(a+b))),\\
\gen 201010A02 ((a+(a+b))+(c+(a+(a+b)))),\\
\gen 2A0001120 ((a+b)+((a+b)+(c+(a+c)))),\\
\gen 2A1011A02 ((a+(a+c))+(c+(a+(a+b)))),\\
\gen A12110201 (b+((a+(a+b))+(c+(a+(a+b))))),\\
\gen 02A100A12 (a+(b+((a+(a+b))+\\*%
&&(c+(a+(a+b)))))),\\
\gen 20A010102 ((a+(a+b))+((a+(a+b))+\\*%
&&(c+(a+c)))),\\
\gen 1A2110A12 (b+(b+((a+(a+b))+\\*%
&&(c+(a+(a+b)))))),\\
\gen 01210002A (a+(a+(b+((a+(a+b))+\\*%
&&(c+(a+(a+b))))))),\\
\gen 0211001A2 (a+(b+(b+((a+(a+b))+\\*%
&&(c+(a+(a+b))))))),\\
\gen AA21A0201 ((b+(a+b))+((a+(a+b))+\\*%
&&(c+(a+(a+b))))),\\
\gen 0A2100021 (a+(a+(b+(b+((a+(a+b))+\\*%
&&(c+(a+(a+b)))))))),\\
\gen 1121A002A ((b+(a+b))+(a+(b+((a+(a+b))+\\*%
                && (c+(a+(a+b))))))),\\
\gen A2A210012 (((a+b)+(c+(a+b)))+\\*%
               &&(a+(a+(b+((a+(a+b))+\\*%
               &&(c+(a+(a+b)))))))),\\
\gen A212100A2 (((a+b)+(c+(a+b)))+\\*%
               &&(a+(a+(b+(b+((a+(a+b))+\\*%
               &&(c+(a+(a+b))))))))),\\
\gen 12A2A0012 (((a+b)+((a+b)+(c+(a+c))))+\\*%
               &&(a+(a+(b+((a+(a+b))+\\*%
               &&(c+(a+(a+b)))))))),\\
\gen A01010A2A ((a+(a+b))+(((a+b)+(c+(a+b)))+\\*%
               &&(a+(a+(b+((a+(a+b))+\\*%
               &&(c+(a+(a+b))))))))),\\
\gen 1212A00A2 (((a+b)+((a+b)+(c+(a+c))))+\\*%
               &&(a+(a+(b+(b+((a+(a+b))+\\*%
               &&(c+(a+(a+b))))))))),\\
\gen 101010A21 ((a+(a+b))+(((a+b)+(c+(a+b)))+\\*%
               &&(a+(a+(b+(b+((a+(a+b))+\\*%
               &&(c+(a+(a+b)))))))))).
\end{eqnarray*}
}

So far, we have studied the implosion of the quantum jigsaw puzzle.
What about its explosion?
What if we try to actually measure the two-valued probability
assignments?

First of all, we have to clarify what ``measurement'' means. Indeed, in
the three--dimensional cases, from all the numerous tripods represented
here as lines, only a single one can actually be ``measured'' in a
straightforward way. All the others have to be counterfactually
inferred.

Thus, of course, only if {\em all} propositions --- and not just the
ones
which are comeasurable --- are counterfactually inferred and compared,
we
would end up in a complete contradiction. In doing so, we accept the EPR
definition of
``element of physical reality.''
As a fall-back option we may be willing
to accept that ``actual elements of physical reality'' are determined
only by the measurement context.

This is not as mindboggling as it first may appear.
It should be noted that in finite--dimensional Hilbert
spaces, any two {\em commuting} self-adjoint operators
$A$ and $B$ corresponding to observables can be
simultaneously diagonalized
\cite[section 79]{halmos-vs}.
Furthermore, $A$ and $B$ commute if and only if there
exists a self-adjoint ``Ur''-operator $U$ and two real-valued functions
\index{Ur-operator}
$f$ and
$g$ such that $A=f(U)$ and $B=g(U)$
(cf.
\cite[Section 84]{halmos-vs},
Varadarajan
\cite[p. 119-120, Theorem 6.9]{varadarajanI} and
  Pt{\'{a}}k and Pulmannov{\'{a}}
\cite[p. 89, Theorem 4.1.7]{pulmannova-91}).
 A generalization to an arbitrary
number of
mutually commuting operators is straightforward. Stated pointedly: every
set of mutually commuting observables
can be represented by just one ``Ur''-operator, such that all the
operators are functions thereof.

One example is the spin one-half case. There, for instance, the
commuting operators are $A={\Bbb I}$ and
$B=\sigma_1$ (uncritical factors have been omitted). In this case,
take
$U=B$ and
$f(x)=x^2$,
$g(x)=x$.

For spin component measurements along the Cartesian coordinate axes
$(1,0,0)$, $(0,1,0)$ and $(0,0,1)$,
the ``Ur''-operator for the tripods used for the construction of the
Kochen-Specker paradox is ($\hbar = 1$)\cite{kochen1}
\begin{equation}
U=aJ_1^2+bJ_2^2+cJ_3^2=
 {1\over 2}
\left(
\begin{array}{ccc}
a+b+2c&0&a-b\\
0&2a+2b&0\\
a-b&0&a+b+2c
\end{array}
\right),
\label{le-ndfuo}
\end{equation}
where $a,b$ and $c$ are mutually distinct real numbers and
$$
J_1^2= {1\over 2}
\left(
\begin{array}{ccc}
1&0&1\\
0&2&0\\
1&0&1
\end{array}
\right), \;
J_2^2= {1\over 2}
\left(
\begin{array}{ccc}
1&0&-1\\
0&2&0\\
-1&0&1
\end{array}
\right), \;
J_3^2=
\left(
\begin{array}{ccc}
1&0&0\\
0&0&0\\
0&0&1\\
\end{array}
\right)
$$
are
the squares of the spin state observables.
Since $J_1^2,J_2^2,J_3^2$ are commuting and all functions of $U$,
they can be identified with the observables constituting the tripods.
(See also References
\cite[pp. 199-200]{peres} and
\cite{rzbb,swift80a}.)

Let us be a little bit more explicit. We have
\begin{eqnarray*}
&&J_1^2=[(a-b)(c-a)]^{-1}[U-(b+c)](U-2a),\\
&&J_2^2=[(a-b)(b-c)]^{-1}[U-(a+c)](U-2b),\\
&&J_3^2=[(c-a)(b-c)]^{-1}[U-(a+b)](U-2c).
\end{eqnarray*}
The diagonal form of the ``Ur''-operator (\ref{le-ndfuo}) is
$$
U=
\left(
\begin{array}{ccc}
a+b&0&0\\
0&b+c&0\\
0&0&a+c
\end{array}
\right).
$$
Measurement of $U$ can, for instance, be realized by a set of beam
splitters
\cite{rzbb}; or in an arrangement proposed by Kochen and Specker
\cite{kochen1}.
Any such measurement will yield either the eigenvalue
$a+b$  (exclusive) or the eigenvalue
$b+c$  (exclusive) or the eigenvalue
$a+c$. Since $a,b,c$ are mutually distinct, one always knows which one
of the eigenvalues it is. Furthermore, we observe that
$$J_1^2+J_2^2+J_3^2=2{\Bbb I}.$$
Since the possible eigenvalues of any $J_i^2, i=1,2,3$ are
either~0 or~1, the eigenvalues of two observables $J_i^2,i=1,2,3$ must
be~1, and one must be~0. Any measurement of the
``Ur''-operator $U$ thus yields
$a+b$ associated with $J_1^2=J_2^2=1$, $J_3^2=0$ (exclusive) or
$a+c$ associated with $J_1^2=J_3^2=1$, $J_2^2=0$ (exclusive) or
$b+c$ associated with $J_2^2=J_3^2=1$, $J_1^2=0$.

We now consider then the following propositions
\begin{description}
\item[$p_1$:] The measurement result of $J_1$ is~0,
\item[$p_2$:] The measurement result of $J_2$ is~0,
\item[$p_3$:] The measurement result of $J_2$ is~0;
\end{description}
or equivalently,
\begin{description}
\item[$p_1$:] The measurement result of $U$ is~$b+c$,
\item[$p_2$:] The measurement result of $U$ is~$a+c$,
\item[$p_3$:] The measurement result of $U$ is~$a+b$.
\end{description}

For spin component measurements along a different set $\bar x,\bar
y,\bar z$ of mutually orthogonal rays,
the ``Ur''-operator is given by
$$\bar U=\bar a{\bar J}_1^2+\bar b{\bar J}_2^2+\bar c{\bar J}_3^2,$$
where
$$
{\bar J}_1=S(\bar x),\quad
{\bar J}_2=S(\bar y),\quad
{\bar J}_3=S(\bar z).$$

Let us, for example, take
$\bar x=(1/\sqrt{2})(1,1,0)$,
$\bar y=(1/\sqrt{2})(-1,1,0)$, and
$\bar z=z$.
In terms of polar coordinates $\theta ,\phi ,r$, these orthogonal
directions are
$\bar x=(\pi /2,\pi /4,1)$,
$\bar y=(\pi /2,-\pi /4,1)$, and
$\bar z=(0,0,1)$, and
\begin{eqnarray*}
{\bar J}_1^2= \left(S({\pi \over 2}, {\pi \over 4})\right)^2&=&
{1\over 2}
\left(
\begin{array}{ccc}
1&0&-i\\
0&2&0\\
i&0&1
\end{array}
\right), \\
{\bar J}_2^2= \left(S({\pi \over 2}, -{\pi \over 4})\right)^2&=&
{1\over 2}
\left(
\begin{array}{ccc}
1&0&i\\
0&2&0\\
-i&0&1
\end{array}
\right), \\
{\bar J}_3^2=\left(S(0,0)\right)^2&=&
\left(
\begin{array}{ccc}
1&0&0\\
0&0&0\\
0&0&1\\
\end{array}
\right).
\end{eqnarray*}
The ``Ur''-operator takes on the matrix form
$$
\bar U=
 {1\over 2}
\left(
\begin{array}{ccc}
{\bar a}+{\bar b}+2{\bar c}&0&-i{\bar a}+i\bar b\\
0&2{\bar a}+2\bar b&0\\
i{\bar a}-i\bar b&0&{\bar a}+{\bar b}+2{\bar c}
\end{array}
\right).$$
As expected, the eigenvalues of
$\bar U$ are
$
\bar a+\bar b,
\bar b+\bar c,
\bar a+\bar c$.
This result holds true for arbitrary
rotations $S0(3)$
\cite{murnaghan}
of the coordinate axes
(tripod), parameterized, for instance, by the Euler angles
$\alpha ,\beta , \gamma$.

Hence, stated pointedly and repeatedly, any measurement of elements of
physical reality boils down, in
a sense, to measuring a {\em single} ``Ur''-observable, from which the
three observables in the tripod can be derived. Different tripods
correspond to different ``Ur''-observables.

\section{Algebraic options for a ``completion'' of quantum mechanics}
Just how far might a classical understanding of quantum mechanics in
principle be possible?  We shall attempt an answer to this question in
terms of mappings of quantum universes into classical ones, more
specifically in terms of embeddings of quantum logics into classical
logics. We shall also shortly discuss surjective extensions (many-to-one
mappings) of classical logics into quantum logics.\footnote{
No attempt will be made here to give a comprehensive review of hidden
parameter models. See, for instance, an article by
Gudder
\cite{gudder1}, where
a different approach to the question of hidden parameters is pursued.
For a historical review, see the books by Jammer
\cite{jammer:66,jammer1}.}

It is always possible to enlarge a quantum logic to a classical
logic, thereby mapping the quantum logic into the classical logic. In
algebraic terms, the question  is how much structure can be preserved.

A possible ``completion'' of quantum mechanics had already been
suggested, though in not very
concrete terms, by Einstein, Podolsky and Rosen (EPR) \cite{epr}. These authors
speculated that ``elements of physical reality'' with definite values
exist irrespective
of whether or not they are actually observed. Moreover, EPR conjectured,
the
quantum formalism can be ``completed'' or ``embedded'' into a larger theoretical framework
which would reproduce the quantum theoretical results but would otherwise
be classical and deterministic from an algebraic and logical point of view.

A proper formalization of the term ``element of physical reality''
suggested by EPR can be given in terms of two-valued states or valuations,
which can take on only one of two values $0$ and $1$ and which are
interpretable as the classical logical truth assignments {\it false} and
{\it true}, respectively.  Recall that Kochen and Specker's results
\cite{kochen1} state that for quantum systems representable by Hilbert
spaces of dimension higher than two, there does not exist any such valuation
$s: L\rightarrow \{0,1\}$
on the set of closed linear subspaces $L$ interpretable as quantum
mechanical propositions preserving the lattice operations
and the orthocomplement,
even if these lattice operations are carried out among commuting
(orthogonal) elements only. Moreover,
the set of truth assignments on quantum logics is not
separating and not unital. That is, there exist  different quantum propositions
which cannot be distinguished by any classical truth assignment.
(For related arguments
and conjectures based upon a theorem by Gleason \cite{Gleason}, see
Zierler and Schlessinger \cite{ZirlSchl-65} and John Bell
\cite{bell-66}.)

Particular emphasis will be given to embeddings of
quantum universes into classical ones which do not
necessarily preserve (binary lattice) operations identifiable with the
logical {\it or} and {\it and} operations. Stated pointedly, if one is
willing to abandon the preservation of quite commonly used logical
functions, then it is possible to give a classical meaning to quantum
physical statements, thus giving rise to an ``understanding'' of
quantum mechanics.

One of the questions already raised in Specker's almost forgotten first
article \cite{specker-60} concerned an embedding of a quantum
logical structure $L$ of propositions into a classical universe
represented by Boolean algebras $B$.  Such an embedding can be
formalized as a function $\varphi :L\rightarrow B$ with the following
properties (Specker had a modified notion of embedding in mind; see
below).  Let $p,q\in L$.

\begin{description}

\item[(i)]
Injectivity:
two different quantum logical propositions are mapped into two
different propositions of the Boolean algebra; i.e., if $p\neq
q $ then $ \varphi (p)\neq \varphi (q)$.

\item[(ii)]
Preservation of the order relation:
if $p\rightarrow q$ then $\varphi (p) \rightarrow \varphi (q)$.

\item[(iii)]
Preservation of the lattice operations, in particular preservation of the
\begin{description}
\item[(ortho-)complement]
$\varphi(p')=\varphi (p)'$;
\item[{\it or} operation] $\varphi
(p\vee q)=\varphi (p) \vee \varphi (q)$;
\item[{\it and} operation] $\varphi
(p\wedge q)=\varphi (p) \wedge \varphi (q)$.
\end{description}
\end{description}

It is rather obvious that we cannot have an embedding from the
quantum to
the classical universe satisfying all three requirements (i)--(iii).  In
particular, a head-on approach requiring (iii) is doomed to failure,
since the nonpreservation of lattice operations among
noncomeasurable propositions is quite evident, given the nondistributive
structure of quantum logics.

One method of embedding any arbitrary partially ordered set into a
concrete orthomodular lattice which in turn can be embedded into a
Boolean algebra has been used by Kalmbach \cite{kalmbach-77} and
extended by Harding \cite{harding-91} and Mayet and Navara
\cite{navara-95}. These
{\it Kalmbach embeddings}, as they may be called, are based upon the
following two theorems.
Given any poset $P$,
 there is an orthomodular lattice $L$ and an
embedding $\varphi :P\rightarrow L=K(P)$ such that if $x,y\in P$, then
(i) $x\le y$ if and only if $\varphi (x)\le \varphi (y)$,
(ii) if $x\wedge y$ exists, then
$\varphi (x)\wedge \varphi (y)=\varphi (x \wedge y)$,    and
(iii) if $x\vee y$ exists, then
$\varphi (x)\vee \varphi (y)=\varphi (x \vee y)$
\cite{kalmbach-77}.\footnote{Kalmbach's original result referred to
an arbitrary lattice instead of the poset $P$, but by the MacNeille
 completion \cite{macneille} it is always possible
to embedd a poset into a lattice, thereby  preserving the order relation
and the meets and joins, if they exist \cite{harding-priv}.
Also, a direct proof has been given by Navara.}
Furthermore, $L$ in the above result has a full set of two-valued states
\cite{harding-91,navara-95} and thus can be embedded into a Boolean
algebra $B$ by preserving lattice operations among orthogonal elements
and additionally by preserving the orthocomplement.

Note that the {\em combined}  Kalmbach embedding $P\rightarrow
K(P) \rightarrow B = P\rightarrow B$ does not necessarily preserve the
logical
{\it and},
{\it
or} and {\it not} operations.
(There may not even be a complement defined on the
partially ordered set which is embedded.)
Nevertheless, every chain of the original poset gets embedded into a
Boolean algebra whose lattice operations are totally preserved under the
combined Kalmbach embedding.

The Kalmbach embedding of some bounded lattice $L$ into a concrete
orthomodular lattice
$K(L)$ may be thought of as the pasting of Boolean algebras
corresponding to all maximal chains of $L$ \cite{harding-priv}.

First, let us consider  linear chains
$0=a_0\rightarrow
a_1\rightarrow a_2\rightarrow \cdots \rightarrow 1=a_m$.
Such chains
generate Boolean algebras $2^{m-1}$ in the following way: from the first
nonzero element $a_1$ on to the greatest element $1$, form
$A_n=a_n\wedge (a_{n-1})'$, where $(a_{n-1})'$ is the complement of $a_{n-1}$
relative to $1$; i.e., $(a_{n-1})'=1-a_{n-1}$.  $A_n$ is then an atom of the
Boolean algebra generated by the bounded chain
$0=a_0\rightarrow
a_1\rightarrow a_2\rightarrow \cdots \rightarrow 1$.

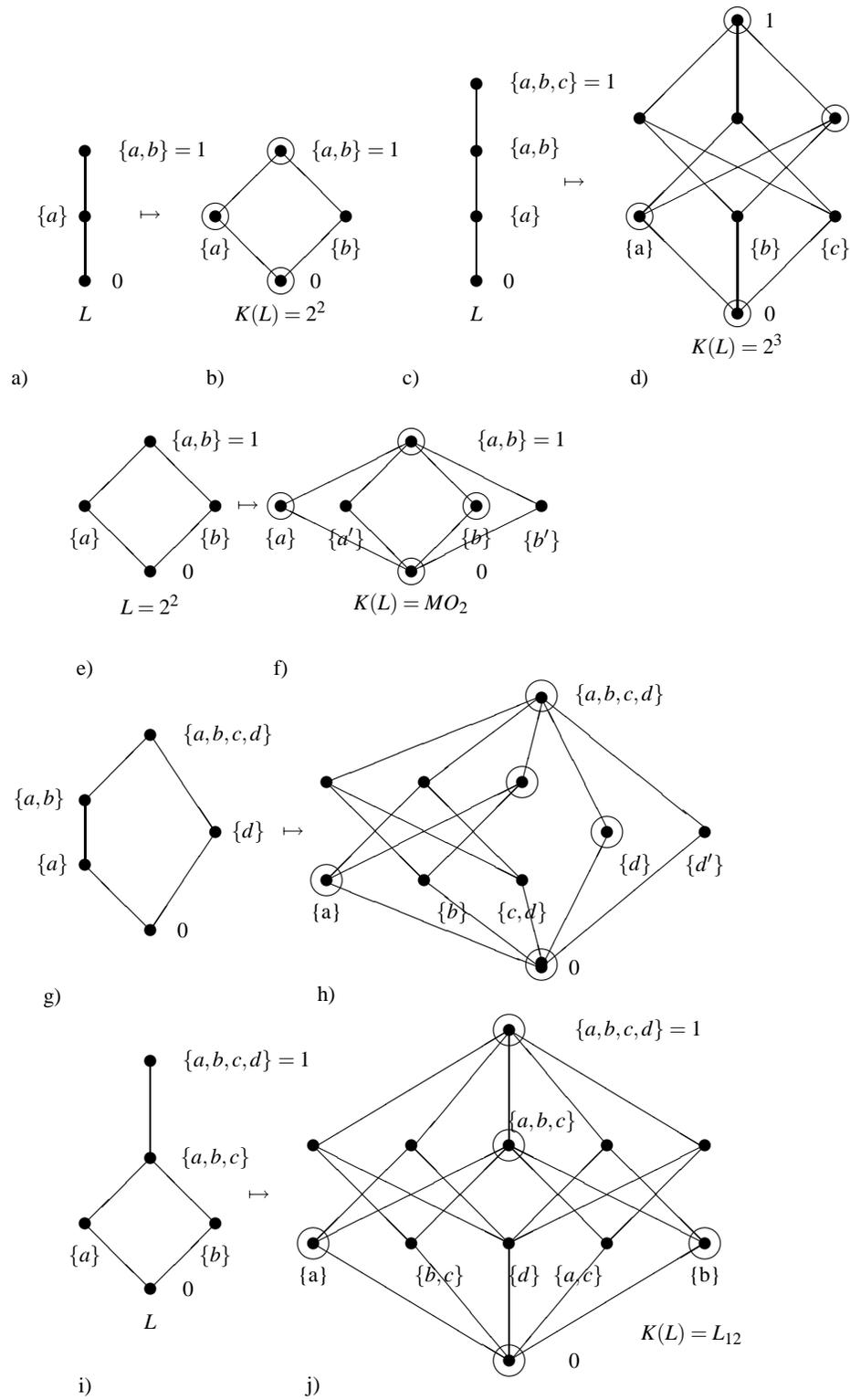
\begin{figure}
\begin{center}
\unitlength 0.95mm
\linethickness{0.4pt}
\begin{picture}(127.00,211.52)
\put(10.00,169.52){\circle*{1.89}}
\put(10.00,169.52){\line(0,1){10.00}}
\put(10.00,179.52){\circle*{1.89}}
\put(10.00,189.52){\circle*{1.89}}
\put(10.00,179.52){\line(0,1){10.00}}
\put(15.00,169.52){\makebox(0,0)[cc]{0}}
\put(5.00,179.52){\makebox(0,0)[cc]{$\{a\}$}}
\put(15.00,189.52){\makebox(0,0)[lc]{$\{a,b\}=1$}}
\put(40.00,169.52){\circle*{1.89}}
\put(40.00,189.52){\circle*{1.89}}
\put(30.00,179.52){\circle*{1.89}}
\put(50.00,179.52){\circle*{1.89}}
\put(50.00,179.52){\line(-1,1){10.00}}
\put(40.00,189.52){\line(-1,-1){10.00}}
\put(30.00,179.52){\line(1,-1){10.00}}
\put(40.00,169.52){\line(1,1){10.00}}
\put(45.33,169.52){\makebox(0,0)[cc]{$0$}}
\put(51.43,189.52){\makebox(0,0)[cc]{$\{a,b\}=1$}}
\put(0.00,154.52){\makebox(0,0)[cc]{a)}}
\put(30.00,154.52){\makebox(0,0)[cc]{b)}}
\put(70.00,169.52){\circle*{1.89}}
\put(70.00,169.52){\line(0,1){10.00}}
\put(70.00,179.52){\circle*{1.89}}
\put(70.00,189.52){\circle*{1.89}}
\put(70.00,179.52){\line(0,1){10.00}}
\put(75.00,169.52){\makebox(0,0)[cc]{0}}
\put(75.00,179.52){\makebox(0,0)[lc]{$\{a\}$}}
\put(60.00,154.52){\makebox(0,0)[cc]{c)}}
\put(70.00,199.85){\circle*{1.89}}
\put(70.00,189.85){\line(0,1){10.00}}
\put(75.00,189.85){\makebox(0,0)[lc]{$\{a,b\}$}}
\put(75.00,199.85){\makebox(0,0)[lc]{$\{a,b,c\}=1$}}
\put(20.00,179.52){\makebox(0,0)[cc]{$\mapsto$}}
\put(85.00,184.52){\makebox(0,0)[cc]{$\mapsto$}}
\put(47.00,77.67){\circle*{1.89}}
\put(62.00,77.67){\circle*{1.89}}
\put(77.00,77.67){\circle*{1.89}}
\put(47.00,92.67){\circle*{1.89}}
\put(62.00,92.67){\circle*{1.89}}
\put(77.00,92.67){\circle*{1.89}}
\put(110.00,209.52){\circle*{1.89}}
\put(110.00,164.52){\circle*{1.89}}
\put(110.00,164.52){\line(1,1){15.00}}
\put(77.00,77.67){\line(-1,1){15.00}}
\put(110.00,194.52){\line(0,1){15.00}}
\put(110.00,209.52){\line(-1,-1){15.00}}
\put(47.00,92.67){\line(1,-1){15.00}}
\put(110.00,179.52){\line(0,-1){15.00}}
\put(110.00,164.52){\line(-1,1){15.00}}
\put(47.00,77.67){\line(2,1){30.00}}
\put(125.00,194.52){\line(-1,1){15.00}}
\put(62.00,92.67){\line(-1,-1){15.00}}
\put(47.00,92.67){\line(2,-1){30.00}}
\put(62.00,77.67){\line(1,1){15.00}}
\put(47.00,72.67){\makebox(0,0)[cc]{\{a\}}}
\put(66.33,72.67){\makebox(0,0)[cc]{$\{b\}$}}
\put(77.00,72.67){\makebox(0,0)[cc]{$\{c,d\}$}}
\put(115.00,164.52){\makebox(0,0)[cc]{$0$}}
\put(115.00,209.52){\makebox(0,0)[cc]{$1$}}
\put(95.00,154.52){\makebox(0,0)[cc]{d)}}
\put(110.00,159.52){\makebox(0,0)[cc]{$K(L)=2^3$}}
\put(20.00,125.00){\circle*{1.89}}
\put(20.00,145.00){\circle*{1.89}}
\put(10.00,135.00){\circle*{1.89}}
\put(30.00,135.00){\circle*{1.89}}
\put(30.00,135.00){\line(-1,1){10.00}}
\put(20.00,145.00){\line(-1,-1){10.00}}
\put(10.00,135.00){\line(1,-1){10.00}}
\put(20.00,125.00){\line(1,1){10.00}}
\put(25.00,125.00){\makebox(0,0)[lc]{0}}
\put(30.00,145.00){\makebox(0,0)[cc]{$\{a,b\}=1$}}
\put(20.00,120.00){\makebox(0,0)[cc]{$L=2^2$}}
\put(10.00,110.00){\makebox(0,0)[cc]{e)}}
\put(30.00,174.52){\makebox(0,0)[cc]{$\{a\}$}}
\put(50.00,174.52){\makebox(0,0)[cc]{$\{b\}$}}
\put(10.00,130.00){\makebox(0,0)[cc]{$\{a\}$}}
\put(30.00,130.00){\makebox(0,0)[cc]{$\{b\}$}}
\put(60.00,125.00){\circle*{1.89}}
\put(60.00,145.00){\circle*{1.89}}
\put(50.00,135.00){\circle*{1.89}}
\put(70.00,135.00){\circle*{1.89}}
\put(70.00,135.00){\line(-1,1){10.00}}
\put(60.00,145.00){\line(-1,-1){10.00}}
\put(50.00,135.00){\line(1,-1){10.00}}
\put(60.00,125.00){\line(1,1){10.00}}
\put(70.00,125.00){\makebox(0,0)[lc]{0}}
\put(70.00,145.00){\makebox(0,0)[lc]{$\{a,b\}=1$}}
\put(50.00,130.00){\makebox(0,0)[cc]{$\{a'\}$}}
\put(70.00,130.00){\makebox(0,0)[cc]{$\{b\}$}}
\put(40.00,135.00){\circle*{1.89}}
\put(80.00,135.00){\circle*{1.89}}
\put(60.00,120.00){\makebox(0,0)[cc]{$K(L)=MO_2$}}
\put(40.00,110.00){\makebox(0,0)[cc]{f)}}
\put(60.00,125.00){\line(-2,1){20.00}}
\put(40.00,135.00){\line(2,1){20.00}}
\put(60.00,145.00){\line(2,-1){20.00}}
\put(80.00,135.00){\line(-2,-1){20.00}}
\put(40.00,130.00){\makebox(0,0)[cc]{$\{a\}$}}
\put(80.00,129.67){\makebox(0,0)[cc]{$\{b'\}$}}
\put(35.00,135.00){\makebox(0,0)[cc]{$\mapsto$}}
\put(10.00,164.52){\makebox(0,0)[cc]{$L$}}
\put(40.00,164.52){\makebox(0,0)[cc]{$K(L)=2^2$}}
\put(70.00,164.52){\makebox(0,0)[cc]{$L$}}
\put(20.00,100.00){\circle*{1.89}}
\put(20.00,70.00){\circle*{1.89}}
\put(10.00,80.00){\circle*{1.89}}
\put(10.00,90.00){\circle*{1.89}}
\put(30.00,85.00){\circle*{1.89}}
\put(20.00,70.00){\line(-1,1){10.00}}
\put(10.00,80.00){\line(0,1){10.00}}
\put(10.00,90.00){\line(1,1){10.00}}
\put(20.00,100.00){\line(2,-3){10.00}}
\put(30.00,85.00){\line(-2,-3){10.00}}
\put(3.00,90.00){\makebox(0,0)[cc]{$\{a,b\}$}}
\put(5.00,80.00){\makebox(0,0)[cc]{$\{a\}$}}
\put(35.00,85.00){\makebox(0,0)[cc]{$\{d\}$}}
\put(25.00,100.00){\makebox(0,0)[lc]{$\{a,b,c,d\}$}}
\put(25.00,70.00){\makebox(0,0)[cc]{$0$}}
\put(95.00,179.52){\circle*{1.89}}
\put(110.00,179.52){\circle*{1.89}}
\put(125.00,179.52){\circle*{1.89}}
\put(95.00,194.52){\circle*{1.89}}
\put(110.00,194.52){\circle*{1.89}}
\put(125.00,194.52){\circle*{1.89}}
\put(125.00,179.52){\line(-1,1){15.00}}
\put(95.00,194.52){\line(1,-1){15.00}}
\put(95.00,179.52){\line(2,1){30.00}}
\put(110.00,194.52){\line(-1,-1){15.00}}
\put(95.00,194.52){\line(2,-1){30.00}}
\put(110.00,179.52){\line(1,1){15.00}}
\put(95.00,174.52){\makebox(0,0)[cc]{\{a\}}}
\put(114.33,174.52){\makebox(0,0)[cc]{$\{b\}$}}
\put(125.00,174.52){\makebox(0,0)[cc]{$\{c\}$}}
\put(90.00,85.00){\circle*{1.89}}
\put(105.00,85.00){\circle*{1.89}}
\put(94.33,80.00){\makebox(0,0)[cc]{$\{d\}$}}
\put(105.00,80.00){\makebox(0,0)[cc]{$\{d'\}$}}
\put(80.00,105.67){\circle*{1.89}}
\put(80.00,65.00){\circle*{1.89}}
\put(47.00,92.67){\line(5,2){33.00}}
\put(80.00,105.87){\line(-4,-3){18.00}}
\put(77.00,92.67){\line(1,4){3.33}}
\put(80.00,64.33){\circle*{1.89}}
\put(47.00,77.33){\line(5,-2){33.00}}
\put(80.00,64.13){\line(-4,3){18.00}}
\put(77.00,77.33){\line(1,-4){3.33}}
\put(80.00,64.33){\line(1,2){10.33}}
\put(90.33,85.00){\line(-1,2){10.33}}
\put(80.00,105.67){\line(5,-4){25.33}}
\put(105.33,85.40){\line(-6,-5){25.33}}
\put(85.00,64.33){\makebox(0,0)[cc]{$0$}}
\put(85.00,106.00){\makebox(0,0)[lc]{$\{a,b,c,d\}$}}
\put(80.00,64.67){\circle{4.85}}
\put(47.00,77.67){\circle{4.85}}
\put(77.00,92.67){\circle{4.85}}
\put(80.00,106.00){\circle{4.85}}
\put(90.00,85.00){\circle{4.85}}
\put(20.00,50.00){\circle*{1.89}}
\put(20.00,15.00){\circle*{1.89}}
\put(20.00,35.00){\circle*{1.89}}
\put(10.00,25.00){\circle*{1.89}}
\put(30.00,25.00){\circle*{1.89}}
\put(30.00,25.00){\line(-1,1){10.00}}
\put(20.00,35.00){\line(-1,-1){10.00}}
\put(10.00,25.00){\line(1,-1){10.00}}
\put(20.00,15.00){\line(1,1){10.00}}
\put(25.00,15.00){\makebox(0,0)[lc]{0}}
\put(30.00,35.00){\makebox(0,0)[cc]{$\{a,b,c\}$}}
\put(20.00,10.00){\makebox(0,0)[cc]{$L$}}
\put(10.00,0.00){\makebox(0,0)[cc]{i)}}
\put(10.00,20.00){\makebox(0,0)[cc]{$\{a\}$}}
\put(30.00,20.00){\makebox(0,0)[cc]{$\{b\}$}}
\put(36.67,29.33){\makebox(0,0)[cc]{$\mapsto$}}
\put(5.00,59.67){\makebox(0,0)[cc]{g)}}
\put(47.00,60.00){\makebox(0,0)[cc]{h)}}
\put(20.00,35.00){\line(0,1){15.00}}
\put(25.00,50.00){\makebox(0,0)[lc]{$\{a,b,c,d\}=1$}}
\put(45.00,22.00){\circle*{1.89}}
\put(60.00,22.00){\circle*{1.89}}
\put(75.00,22.00){\circle*{1.89}}
\put(45.00,37.00){\circle*{1.89}}
\put(60.00,37.00){\circle*{1.89}}
\put(75.00,37.00){\circle*{1.89}}
\put(75.00,22.00){\line(-1,1){15.00}}
\put(45.00,37.00){\line(1,-1){15.00}}
\put(45.00,22.00){\line(2,1){30.00}}
\put(60.00,37.00){\line(-1,-1){15.00}}
\put(45.00,37.00){\line(2,-1){30.00}}
\put(60.00,22.00){\line(1,1){15.00}}
\put(45.00,17.00){\makebox(0,0)[cc]{\{a\}}}
\put(64.33,17.00){\makebox(0,0)[cc]{$\{b,c\}$}}
\put(77.33,17.00){\makebox(0,0)[cc]{$\{d\}$}}
\put(45.00,22.00){\circle{4.85}}
\put(75.00,37.00){\circle{4.85}}
\put(105.00,22.00){\circle*{1.89}}
\put(90.00,22.00){\circle*{1.89}}
\put(105.00,37.00){\circle*{1.89}}
\put(90.00,37.00){\circle*{1.89}}
\put(75.00,22.00){\line(1,1){15.00}}
\put(105.00,37.00){\line(-1,-1){15.00}}
\put(105.00,22.00){\line(-2,1){30.00}}
\put(90.00,37.00){\line(1,-1){15.00}}
\put(105.00,37.00){\line(-2,-1){30.00}}
\put(90.00,22.00){\line(-1,1){15.00}}
\put(105.00,17.00){\makebox(0,0)[cc]{\{b\}}}
\put(85.67,17.00){\makebox(0,0)[cc]{$\{a,c\}$}}
\put(105.00,22.00){\circle{4.85}}
\put(45.00,36.67){\line(5,3){30.00}}
\put(75.00,54.67){\line(5,-3){30.00}}
\put(90.00,37.00){\line(-5,6){14.72}}
\put(75.28,54.67){\line(-5,-6){14.72}}
\put(75.00,37.00){\line(0,1){17.33}}
\put(75.00,54.67){\circle*{1.89}}
\put(75.00,54.67){\circle{4.85}}
\put(45.00,22.00){\line(5,-3){30.00}}
\put(75.00,4.00){\line(5,3){30.00}}
\put(90.00,21.67){\line(-5,-6){14.72}}
\put(75.28,4.00){\line(-5,6){14.72}}
\put(75.00,21.67){\line(0,-1){17.33}}
\put(75.00,4.00){\circle*{1.89}}
\put(75.00,4.00){\circle{4.85}}
\put(45.00,0.00){\makebox(0,0)[cc]{j)}}
\put(85.00,4.00){\makebox(0,0)[cc]{$0$}}
\put(85.00,54.67){\makebox(0,0)[lc]{$\{a,b,c,d\}=1$}}
\put(80.00,40.67){\makebox(0,0)[cc]{$\{a,b,c\}$}}
\put(40.00,169.52){\circle{4.00}}
\put(30.00,179.52){\circle{4.00}}
\put(40.00,189.52){\circle{4.00}}
\put(110.00,164.52){\circle{4.00}}
\put(95.00,179.52){\circle{4.00}}
\put(125.00,194.52){\circle{4.00}}
\put(110.00,209.52){\circle{4.00}}
\put(60.00,145.00){\circle{4.00}}
\put(40.00,135.00){\circle{4.00}}
\put(70.00,135.00){\circle{4.00}}
\put(60.00,125.00){\circle{4.00}}
\put(103.00,8.00){\makebox(0,0)[cc]{$K(L)=L_{12}$}}
\put(41.90,85.00){\makebox(0,0)[cc]{$\mapsto$}}
\end{picture}
\end{center}
\caption{\label{f-thech}
Examples of Kalmbach embeddings.}
\end{figure}

Take, for example, a three-element chain
$0= a_0\rightarrow \{a\}\equiv a_1\rightarrow
\{a,b\}\equiv 1=a_2$
as depicted in Figure
\ref{f-thech}a).
In this case,
\begin{eqnarray*}
A_1&=&a_1\wedge (a_0)'=a_1\wedge 1\equiv \{a\}\wedge \{a,b\}=\{a\},\\
A_2&=&a_2\wedge (a_1)'=1\wedge (a_1)'\equiv \{a,b\}\wedge \{b\}=\{b\}.
\end{eqnarray*}
This construction results in a four-element Boolean Kalmbach lattice
$K(L)=2^2$ with the two atoms $\{a\}$ and $\{b\}$
depicted in Figure
\ref{f-thech}b).

Take, as a second example, a four-element chain
$0= a_0\rightarrow \{a\}\equiv a_1\rightarrow \{a,b\}
\rightarrow \{a,b,c\}\equiv 1=a_3$
as depicted in Figure
\ref{f-thech}c).
In this case,
\begin{eqnarray*}
A_1&=&a_1\wedge (a_0)'=a_1\wedge 1\equiv \{a\}\wedge \{a,b,c\}=\{a\},\\
A_2&=&a_2\wedge (a_1)'\equiv \{a,b\}\wedge \{b,c\}=\{b\},\\
A_3&=&a_3\wedge (a_2)'=1\wedge (a_2)'\equiv \{a,b,c\}\wedge \{c\}=\{c\}.\\
\end{eqnarray*}
This construction results in a eight-element Boolean Kalmbach lattice
$K(L)=2^3$ with the three atoms $\{a\}$, $\{b\}$ and $\{c\}$
depicted in Figure
\ref{f-thech}d).

To apply Kalmbach's
construction to any bounded lattice, all Boolean
algebras generated by the maximal chains of the lattice are pasted
together.
An element common to
two or more maximal chains must be common to the blocks they generate.

Take, as a third example, the Boolean lattice $2^2$ drawn in Figure
\ref{f-thech}e).
$2^2$ contains two
linear chains of length three which are pasted together horizontally at
their smallest and biggest elements.  The resulting
Kalmbach lattice $K(2^2)=MO_2$ is of the ``Chinese lantern'' type, as
depicted in Figure
\ref{f-thech}f).

Take, as a fourth example, the pentagon drawn in Figure
\ref{f-thech}g).
It contains two
linear chains. One is of length three, the other is of length four. The
resulting Boolean algebras $2^2$ and $2^3$ are again horizontally pasted
together at their extremities $0,1$.
The resulting
Kalmbach lattice is
depicted in Figure
\ref{f-thech}h).

In the
fifth example drawn in Figure
\ref{f-thech}i), the lattice has two maximal chains
which
share a common element.  This element is common to the two Boolean
algebras; and hence central in $K(L)$.
The construction of the five atoms proceeds as follows.
\begin{eqnarray*}
A_1&=& \{a\}\wedge \{a,b,c,d\}=\{a\},\\
A_2&=& \{a,b,c\}\wedge \{b,c,d\}=\{b,c\},\\
A_3&=&B_3= \{a,b,c,d\}\wedge \{d\}=\{d\},\\
B_1&=& \{b\}\wedge \{a,b,c,d\}=\{b\},\\
B_2&=& \{a,b,c\}\wedge \{a,c,d\}=\{a,c\},\\
\end{eqnarray*}
where the two sets of atoms
$\{A_1,A_2,A_3=B_3\}$ and
$\{B_1,B_2,B_3=A_3\}$ span two Boolean algebras $2^3$ pasted together at
the extremities and at $A_3=B_3$ and $A_3'=B_3'$.
 The resulting lattice is $2\times
MO_2=L_{12}$ depicted in Figure
\ref{f-thech}j).

Notice that there is an equivalence of the lattices $K(L)$ resulting
from Kalmbach embeddings and automata partition logics \cite{svozil-93}.
The Boolean subalgebras resulting from maximal chains in
the Kalmbach embedding case correspond to the Boolean subalgebras from
individual automaton experiments. In both cases, these blocks are pasted
together similarly.

\clearpage
\ifx\undefined\bysame
\newcommand{\bysame}{\leavevmode\hbox to3em{\hrulefill}\,}
\fi


\begin{thebibliography}{10}

\bibitem{birkhoff-36}
G. Birkhoff and J. von Neumann, ``The Logic of Quantum Mechanics,'' Annals of
  Mathematics {\bf 37,} 823--843 (1936).

\bibitem{bridgman27}
P.~W. Bridgman, {\em The Logic of Modern Physics} (New York, 1927).

\bibitem{bridgman}
P.~W. Bridgman, ``A Physicist's Second Reaction to {M}engenlehre,'' Scripta
  Mathematica {\bf 2,} 101--117, 224--234 (1934), cf. R. Landauer
  \cite{landauer-95}.

\bibitem{bridgman52}
P.~W. Bridgman, {\em The Nature of Some of Our Physical Concepts}
  (Philosophical Library, New York, 1952).

\bibitem{ma-57}
G.~W. Mackey, ``Quantum mechanics and {H}ilbert space,'' Amer. Math. Monthly,
  Supplement {\bf 64,} 45--57 (1957).

\bibitem{jauch}
J.~M. Jauch, {\em Foundations of Quantum Mechanics} (Addison-Wesley, Reading,
  MA., 1968).

\bibitem{varadarajanI}
V. Varadarajan, {\em Geometry of Quantum Theory I} (van Nostrand, Princeton,
  1968).

\bibitem{varadarajanII}
V.~S. Varadarajan, {\em Geometry of Quantum Theory II} (van Nostrand,
  Princeton, 1970).

\bibitem{piron-76}
C. Piron, {\em Foundations of Quantum Physics} (W. A. Benjamin, Reading, MA,
  1976).

\bibitem{marlow}
A.~R. Marlow, {\em Mathematical Foundations of Quantum Theory} (Academic Press,
  New York, 1978).

\bibitem{gudder:79}
S.~P. Gudder, {\em Stochastic Methods in Quantum Mechanics} (North Holland, New
  York, 1979).

\bibitem{gudder}
S.~P. Gudder, {\em Quantum Probability} (Academic Press, San Diego, 1988).

\bibitem{maczy}
M. Maczy\'nski, ``On a functional representation of the lattice of projections
  on a Hilbert space,'' Studia Math. {\bf 47,} 253--259 (1973).

\bibitem{bell-cas}
E.~G. Beltrametti and G. Cassinelli, {\em The Logic of Quantum Mechanics}
  (Addison-Wesley, Reading, MA, 1981).

\bibitem{kalmbach-83}
G. Kalmbach, {\em Orthomodular Lattices} (Academic Press, New York, 1983).

\bibitem{kalmbach-86}
G. Kalmbach, {\em Measures and Hilbert Lattices} (World Scientific, Singapore,
  1986).

\bibitem{cohen}
D.~W. Cohen, {\em An Introduction to Hilbert Space and Quantum Logic}
  (Springer, New York, 1989).

\bibitem{pulmannova-91}
P. Pt{\'{a}}k and S. Pulmannov{\'{a}}, {\em Orthomodular Structures as Quantum
  Logics} (Kluwer Academic Publishers, Dordrecht, 1991).

\bibitem{giuntini-91}
R. Giuntini, {\em Quantum Logic and Hidden Variables} (BI Wissenschaftsverlag,
  Mannheim, 1991).

\bibitem{svozil-ql}
K. Svozil, {\em Quantum Logic} (Springer, Singapore, 1998).

\bibitem{pavicic-92}
M. Pavi{\v{c}}i{\'{c}}, ``Bibliography on Quantum Logics and Related
  Structures,'' International Journal of Theoretical Physics {\bf 31,} 373--461
  (1992).

\bibitem{piz-88}
R. Piziak, ``Orthomodular lattices and quadratic spaces: a survey,'' Rocky
  Mountain Journal of Mathematics {\bf 21,} 951--992 (1991).

\bibitem{nav:91}
M. Navara and V. Rogalewicz, ``The pasting constructions for orthomodular
  posets,'' Mathematische Nachrichten {\bf 154,} 157--168 (1991).

\bibitem{kalmbach-priv}
G. Kalmbach, private communication (unpublished).

\bibitem{svozil-93}
K. Svozil, {\em Randomness \& Undecidability in Physics} (World Scientific,
  Singapore, 1993).

\bibitem{harding-priv}
J. Harding, private communication, March 1998 (unpublished).

\bibitem{cheval-priv}
G. Chevalier, private communication, March 1998 (unpublished).

\bibitem{cheval-or}
G. Chevalier, ``Commutators and decompositions of orthomodular lattices,''
  Order {\bf 6,} 181--194 (1989).

\bibitem{svozil-unev}
K. Svozil, ``Undecidability everywhere?,''  in {\em Boundaries and Barriers. On
  the Limits to Scientific Knowledge}, J.~L. Casti and A. Karlquist, eds.,
  (Addison-Wesley, Reading, MA, 1996), \ pp.\ 215--237.

\bibitem{Boole}
G. Boole, {\em An investigation of the laws of thought} (Dover edition, New
  York, 1958).

\bibitem{Boole-62}
G. Boole, ``On the theory of probabilities,'' Philosophical Transactions of the
  Royal Society of London {\bf 152,} 225--252 (1862).

\bibitem{froissart-81}
M. Froissart, ``Constructive generalization of {B}ell's inequalities,'' Nuovo
  Cimento B {\bf 64,} 241--251 (1981).

\bibitem{cirelson:80}
B.~S. {Cirel'son (=Tsirel'son)}, ``Quantum generalizations of {B}ell's
  inequality,'' Letters in Mathematical Physics {\bf 4,} 93--100 (1980).

\bibitem{cirelson}
B.~S. {Cirel'son (=Tsirel'son)}, ``Some results and problems on quantum
  {B}ell-type inequalities,'' Hadronic Journal Supplement {\bf 8,} 329--345
  (1993).

\bibitem{pitowsky-86}
I. Pitowsky, ``The range of quantum probabilities,'' J. Math. Phys. {\bf 27,}
  1556--1565 (1986).

\bibitem{pitowsky}
I. Pitowsky, {\em Quantum Probability---Quantum Logic} (Springer, Berlin,
  1989).

\bibitem{pitowsky-89a}
I. Pitowsky, ``From {G}eorge {B}oole to {J}ohn {B}ell: The origin of {B}ell's
  inequality,''  in {\em {B}ell's Theorem, Quantum Theory and the Conceptions
  of the Universe}, M. Kafatos, ed., (Kluwer, Dordrecht, 1989), \ pp.\ 37--49.

\bibitem{Pit-91}
I. Pitowsky, ``Correlation polytopes their geometry and complexity,''
  Mathematical Programming {\bf 50,} 395--414 (1991).

\bibitem{Pit-94}
I. Pitowsky, ``{G}eorge {B}oole's `Conditions od Possible Experience' and the
  Quantum Puzzle,'' Brit. J. Phil. Sci. {\bf 45,} 95--125 (1994).

\bibitem{ziegler}
G.~M. Ziegler, {\em Lectures on Polytopes} (Springer, New York, 1994).

\bibitem{2000-poly}
I. Pitowsky and K. Svozil, ``New optimal tests of quantum nonlocality,''
  Physical Review A {\bf 64,} 014102 (2001).

\bibitem{2001-cddif}
S. Filipp and K. Svozil, ``{B}oole-{B}ell-type inequalities in Mathematica,''
  In {\em Challenging the Boundaries of Symbolic Computation, Proceedings of
  the 5th International {M}athematica Symposium}, P. Mitic, P. Ramsden, and J.
  Carne, eds.,\ pp.\ 215--222  (Imperial College Press, London, 2003).

\bibitem{garey}
M.~R. Garey and D.~S. Johnson, {\em Computers and Intractability. {A} Guide to
  the Theory of {NP}-Completeness} (Freeman, San Francisco, 1979).

\bibitem{schaller-96}
M. Schaller and K. Svozil, ``Automaton logic,'' International Journal of
  Theoretical Physics {\bf 35,} 911--940 (1996).

\bibitem{dvur-pul-svo}
A. Dvure{\v{c}}enskij, S. Pulmannov{\'{a}}, and K. Svozil, ``Partition Logics,
  Orthoalgebras and Automata,'' Helvetica Physica Acta {\bf 68,} 407--428
  (1995).

\bibitem{cal-sv-yu}
C. Calude, E. Calude, K. Svozil, and S. Yu, ``Physical versus Computational
  Complementarity {I},'' International Journal of Theoretical Physics {\bf 36,}
  1495--1523 (1997).

\bibitem{wright:pent}
R. Wright, ``The state of the pentagon. {A} nonclassical example,''  in {\em
  Mathematical Foundations of Quantum Theory}, A.~R. Marlow, ed., (Academic
  Press, New York, 1978), \ pp.\ 255--274.

\bibitem{wright}
R. Wright, ``Generalized urn models,'' Foundations of Physics {\bf 20,}
  881--903 (1990).

\bibitem{Gleason}
A.~M. Gleason, ``Measures on the closed subspaces of a {H}ilbert space,''
  Journal of Mathematics and Mechanics {\bf 6,} 885--893 (1957).

\bibitem{specker-60}
E. Specker, ``{D}ie {L}ogik nicht gleichzeitig entscheidbarer {A}ussagen,''
  Dialectica {\bf 14,} 175--182 (1960), reprinted in \cite[pp.
  175--182]{specker-ges}; {E}nglish translation: {\it The logic of propositions
  which are not simultaneously decidable}, reprinted in \cite[pp.
  135-140]{hooker}.

\bibitem{specker-ges}
E. Specker, {\em Selecta} (Birkh{\"{a}}user Verlag, Basel, 1990).

\bibitem{schrodinger}
E. Schr{\"{o}}dinger, ``Die gegenw{\"{a}}rtige {S}ituation in der
  {Q}uantenmechanik,'' Naturwissenschaften {\bf 23,} 807--812, 823--828,
  844--849 (1935), {E}nglish translation in \cite{trimmer} and \cite[pp.
  152-167]{wheeler-Zurek:83}; http://www.emr.hibu.no/lars/eng/cat/.

\bibitem{bell-66}
J.~S. Bell, ``On the Problem of hidden variables in quantum mechanics,''
  Reviews of Modern Physics {\bf 38,} 447--452 (1966), reprinted in \cite[pp.
  1-13]{bell-87}.

\bibitem{redhead}
M. Redhead, {\em Incompleteness, Nonlocality, and Realism: A Prolegomenon to
  the Philosophy of Quantum Mechanics} (Clarendon Press, Oxford, 1990).

\bibitem{peres-91}
A. Peres, ``Two simple proofs of the {K}ochen--{S}pecker theorem,'' Journal of
  Physics {\bf A24,} L175--L178 (1991), cf. \cite[pp. 186-200]{peres}.

\bibitem{stairs83}
A. Stairs, ``Quantum logic, realism, and value definiteness,'' Philosophy of
  Science {\bf 50,} 578--602 (1983).

\bibitem{jammer-92}
M. Jammer, ``John {S}teward {B}ell and the debate on the significance of his
  contributions to the foundations of quantum mechanics,''  in {\em Bell's
  Theorem and the Foundations of Modern Physics}, A. van~der Merwe, F. Selleri,
  and G. Tarozzi, eds., (World Scientific, Singapore, 1992), \ pp.\ 1--23.

\bibitem{brown}
H.~R. Brown, ``Bell's other theorem and its connection with nonlocality, part
  1,'' In {\em Bell's {T}heorem and the {F}oundations of {M}odern {P}hysics},
  A. van~der Merwe, F. Selleri, and G. Tarozzi, eds.,\ pp.\ 104--116  (World
  Scientific, Singapore, 1992).

\bibitem{peres}
A. Peres, {\em Quantum Theory: Concepts and Methods} (Kluwer Academic
  Publishers, Dordrecht, 1993).

\bibitem{penrose-ks}
J. Zimba and R. Penrose, ``On {B}ell non-locality without probabilities: more
  curious geometry,'' Studies in History and Philosophy of Modern Physics {\bf
  24,} 697--720 (1993).

\bibitem{clifton-93}
R. Clifton, ``Getting contextual and nonlocal elements--of--reality the easy
  way,'' American Journal of Physics {\bf 61,} 443--447 (1993).

\bibitem{mermin-93}
N.~D. Mermin, ``Hidden variables and the two theorems of {J}ohn {B}ell,''
  Reviews of Modern Physics {\bf 65,} 803--815 (1993).

\bibitem{svozil-tkadlec}
K. Svozil and J. Tkadlec, ``Greechie diagrams, nonexistence of measures in
  quantum logics and {K}ochen--{S}pecker type constructions,'' Journal of
  Mathematical Physics {\bf 37,} 5380--5401 (1996).

\bibitem{havlicek}
H. Havlicek and K. Svozil, ``Density conditions for quantum propositions,''
  Journal of Mathematical Physics {\bf 37,} 5337--5341 (1996).

\bibitem{tkadlec-96}
J. Tkadlec, ``Greechie diagrams of small quantum logics with small state
  spaces,'' International Journal of Theoretical Physics {\bf 37,} 203--209
  (1998).

\bibitem{halmos-vs}
P.~R. Halmos, {\em Finite-dimensional vector spaces} (Springer, New York,
  Heidelberg, Berlin, 1974).

\bibitem{kochen1}
S. Kochen and E.~P. Specker, ``The Problem of Hidden Variables in Quantum
  Mechanics,'' Journal of Mathematics and Mechanics {\bf 17,} 59--87 (1967),
  reprinted in \cite[pp. 235--263]{specker-ges}.

\bibitem{rzbb}
M. Reck, A. Zeilinger, H.~J. Bernstein, and P. Bertani, ``Experimental
  realization of any discrete unitary operator,'' Physical Review Letters {\bf
  73,} 58--61 (1994).

\bibitem{swift80a}
A.~R. Swift and R. Wright, ``Generalized {S}tern--{G}erlach experiments and the
  observability of arbitrary spin operators,'' Journal of Mathematical Physics
  {\bf 21,} 77--82 (1980).

\bibitem{murnaghan}
F.~D. Murnaghan, {\em The Unitary and Rotation Groups} (Spartan Books,
  Washington, D.C., 1962).

\bibitem{gudder1}
S.~P. Gudder, ``On hidden-variable theories,'' Journal of Mathematical Physics
  {\bf 11,} 431--436 (1970).

\bibitem{jammer:66}
M. Jammer, {\em The Conceptual Development of Quantum Mechanics} (McGraw-Hill
  Book Company, New York, 1966).

\bibitem{jammer1}
M. Jammer, {\em The Philosophy of Quantum Mechanics} (John Wiley \& Sons, New
  York, 1974).

\bibitem{epr}
A. Einstein, B. Podolsky, and N. Rosen, ``Can quantum-mechanical description of
  physical reality be considered complete?,'' Physical Review {\bf 47,}
  777--780 (1935).

\bibitem{ZirlSchl-65}
N. Zierler and M. Schlessinger, ``Boolean embeddings of orthomodular sets and
  quantum logic,'' Duke Mathematical Journal {\bf 32,} 251--262 (1965).

\bibitem{kalmbach-77}
G. Kalmbach, ``Orthomodular lattices do not satisfy any special lattice
  equation,'' Archiv der Mathematik {\bf 28,} 7--8 (1977).

\bibitem{harding-91}
J. Harding, ``Orthomodular Lattices whose {M}ac{N}eille completions are not
  orthomodular,'' Order {\bf 8,} 93--103 (1991).

\bibitem{navara-95}
R. Mayet and M. Navara, ``Classes of logics representable as kernels of
  measures,''  in {\em Current Issues in Quantum Logic}, G. Pilz, ed.,
  (Teubner, Stuttgart, Wien, 1995), \ pp.\ 241--248.

\bibitem{macneille}
H.~M. MacNeille, ``Partially ordered sets,'' Trans. Amer. Math. Soc. {\bf 42,}
  416--460 (1937).

\bibitem{landauer-95}
R. Landauer, ``Advertisement For a Paper {I} Like,''  in {\em On Limits}, J.~L.
  Casti and J.~F. Traub, eds., (Santa Fe Institute Report 94-10-056, Santa Fe,
  NM, 1994), \ p.\ 39.

\bibitem{hooker}
C.~A. Hooker, {\em The Logico-Algebraic Approach to Quantum Mechanics. {V}olume
  {I}: Historical Evolution} (Reidel, Dordrecht, 1975).

\bibitem{trimmer}
J.~D. Trimmer, ``The present situation in quantum mechanics: a translation of
  {S}chr{\"{o}}dinger's ``cat paradox'','' Proc. Am. Phil. Soc. {\bf 124,}
  323--338 (1980), reprinted in \cite[pp. 152-167]{wheeler-Zurek:83}.

\bibitem{wheeler-Zurek:83}
J.~A. Wheeler and W.~H. Zurek, {\em Quantum Theory and Measurement} (Princeton
  University Press, Princeton, 1983).

\bibitem{bell-87}
J.~S. Bell, {\em Speakable and Unspeakable in Quantum Mechanics} (Cambridge
  University Press, Cambridge, 1987).

\end{thebibliography}

\end{document}